\documentstyle[12pt,epsf]{article}
\newcommand{\be}{\begin{equation}}
\newcommand{\bea}{\begin{eqnarray}}
\newcommand{\eea}{\end{eqnarray}}
\newcommand{\beas}{\begin{eqnarray*}}
\newcommand{\eeas}{\end{eqnarray*}}
\newcommand{\ba}{\begin{array}}
\newcommand{\ea}{\end{array}}
\newcommand{\ee}{\end{equation}}

\newcommand{\tr}{{\rm Tr}\ }

\newcommand{\nbox}{{\,\lower0.9pt\vbox{\hrule \hbox{\vrule height 0.2 cm \hskip
0.2 cm \vrule height 0.2 cm}\hrule}\,}}

\newcommand{\pa}{\partial}

\newcommand\text[1]{\rm #1}

\newcommand{\CC}{\hbox{\xiiss C\kern-.4emI}}
\newcommand{\RR}{\hbox{\xiiss R\kern-.45emI}}
\newcommand{\ZZ}{\hbox{\xiiss Z\kern-.4emZ}}
\newcommand{\CCs}{\hbox{\ixss C\kern-.4emI}}
\newcommand{\ZZs}{\hbox{\ixss Z\kern-.4emZ}}
\newcommand{\pasl}{\pa\kern-.55em /}
\def\href#1#2{#2}

\begin{document}
\begin{titlepage}
\hfill
\vbox{
    \halign{#\hfil         \cr
           hep-th/0207050 \cr
           SU-ITP-02/29 \cr
           } 
      }  
\vspace*{20mm}
\begin{center}
{\Large {\bf Protected Multiplets of M-theory on a Plane Wave}\\ }

\vspace*{15mm}
\vspace*{1mm}
{K. Dasgupta, M. M. Sheikh-Jabbari, M. Van Raamsdonk} 

\vspace*{1cm}

{Department of Physics, Stanford University\\
382 via Pueblo Mall, Stanford CA 94305-4060, USA\\
{\it keshav, jabbari, mav @itp.stanford.edu} }

\vspace*{1cm}
\end{center}

\begin{abstract}
We show that the symmetry algebra governing the interacting part of the matrix 
model for M-theory on the maximally supersymmetric pp-wave is the basic 
classical Lie superalgebra $SU(4|2)$. We determine the $SU(4|2)$ multiplets 
present in the exact spectrum in the limit where $\mu$ (the mass parameter) 
becomes infinite, and find that these include infinitely many BPS multiplets.
Using the representation theory of $SU(4|2)$, we demonstrate that some of these 
BPS multiplets, including all of the vacuum states of the matrix model plus 
certain 
infinite towers of excited states, have energies which are exactly protected 
non-perturbatively for any value of $\mu > 0$. In the large $N$ limit,
these lead to exact quantum states of M-theory on the pp-wave. 
We also show explicitly that there are certain BPS multiplets which do
receive energy corrections by combining with other BPS multiplets to
form ordinary multiplets. 
\end{abstract}

\end{titlepage}

\tableofcontents
\vskip 1cm
\section{Introduction}

In this paper, we continue our analysis \cite{dsv} of the matrix model proposed 
in \cite{bmn} to describe M-theory on the maximally supersymmetric pp-wave 
\cite{penrose,guven,kg,blau}, with Hamiltonian
\bea\label{PPmatrix}
H &=& R\ \tr \left( {1 \over 2} \Pi_A^2 - {1 \over 4} [X_A, X_B]^2 
- {1 \over 2} \Psi^\top \gamma^A [X_A, \Psi] \right)\cr
&& + {R \over 2} \tr 
\Big(\sum_{i=1}^3 {\left({\mu\over 
3R}\right)^2 } X_i^2 + \sum_{a=4}^9 \left({\mu \over 6R}\right)^2 X_a^2 
\cr
&& \qquad \qquad + 
i {\mu \over 4R} \Psi^\top \gamma^{123} \Psi  + i {2\mu \over 3R} 
\epsilon^{ijk} X_i X_j X_k \Big)\ .   
\eea
In our previous work, we noted that for large $\mu$, the matrix model may be 
expanded about each of its classical supersymmetric vacua (corresponding to 
collections of fuzzy-sphere giant gravitons) to give a quadratic action with 
interactions suppressed by powers of $1 \over \mu$. In the limit $\mu = \infty$,
the theory becomes free and one can explicitly diagonalize the Hamiltonian to 
determine the exact spectrum about each of the vacua. Through explicit 
perturbative calculations we then estimated the ranges of parameters and 
energies for which perturbation theory is valid. 

As we noted in \cite{dsv}, an intriguing feature of the matrix model is the 
unusual superalgebra, in which the Hamiltonian does not commute with the 
supersymmetry generators and the anticommutator of supersymmetry generators 
yields rotation generators as well as the Hamiltonian. This latter property 
opens up the possibility of BPS states in the spectrum carrying angular 
momenta. By examining the $\mu = \infty$ spectrum, we found that such BPS states 
are indeed present, and in fact there are infinite towers of BPS states 
annihilated by 2, 4, 6, or 8 supercharges. 

The motivation for the present work is to determine which of these BPS states 
remain BPS (and therefore have protected energies) away from $\mu = \infty$. 
Typically, BPS multiplets may receive energy corrections only by combining with 
other BPS multiplets to form non-BPS multiplets. We wish to understand when this 
can happen in the present case, and therefore determine which multiplets, if 
any, are protected as we vary $\mu$ to finite values. By demonstrating the 
existence of protected multiplets, we will be able to obtain non-trivial 
information about the spectrum beyond the perturbative regime, and even in the 
M-theory limit. 

We now give a concise summary of our results as we outline the paper.

In section 2, we review the symmetry algebra of the model. We show that the 
symmetry algebra of the $SU(N)$ theory is a ``basic classical Lie 
superalgebra'' 
known as $SU(4|2)$ whose bosonic generators are the $SO(3)$ and $SO(6)$ rotation 
generators and the Hamiltonian. Since the Hamiltonian is one of the generators, 
the spectrum of the matrix model at given values of the parameters $N$ and 
$\mu$ is completely determined by which $SU(4|2)$ representations are present. 
In particular, the energy of a representation (measured in units of $\mu$) may 
shift as we vary the parameters only if there exist nearby representations (or 
combinations of representations) with the same $SO(6) \times SO(3)$ state 
content but different energy.

The representation theory of $SU(4|2)$ has been studied by various 
authors including Kac \cite{kac} (who originally classified the basic classical 
Lie 
superalgebras) and Bars et. al. \cite{bars} (who introduced a convenient 
supertableaux notation which we use heavily.) In section 3, we review relevant 
aspects of this representation theory, and describe the complete set of finite-
dimensional, 
positive-energy, unitary representations of $SU(4|2)$, which we display in 
Figure \ref{reps}. We 
note that generic representations (called ``typical'') fall into one-parameter 
families of representations with identical $SO(3) \times SO(6)$ state content 
but energies which vary as a function of the parameter. In addition, there are a 
discrete set of ``atypical'' representations for which no nearby representations 
with the same $SO(3) \times SO(6)$ states but different energy exist. 

In section 4, we describe the physical implications of this representation 
theory. We note that typical representations are free to have energies which 
vary as function of $\mu$ while atypical multiplets cannot receive energy 
corrections unless they combine with other atypical multiplets to form typical 
multiplets. We then determine all possible sets of atypical multiplets which may 
combine (these sets always involve only two multiplets). For a given atypical 
multiplet, there are at most two other atypical multiplets with which it may 
combine. If neither of these complementary multiplets is in the spectrum 
for  
some $\mu = \mu_0$, we may conclude that the energy of the original atypical 
multiplet is protected as we vary $\mu$ to nearby values. We find 
certain special multiplets (including those known as ``doubly
atypical'') which cannot combine with any other multiplets to form a typical 
multiplet. Any such multiplets in the spectrum at some value of $\mu$ must be in 
the spectrum for all values of $\mu$ and the states in these multiplets have 
non-perturbatively protected energies. Finally, we demonstrate the
exisitence of an infinite family of supersymmetric ``indices'' (linear
combinations of occupation numbers for finite collections of
atypical multiplets given  in Eq.(\ref{index})) 
which are exactly protected for all 
values of $\mu>0$. 

We then proceed to apply this knowledge to the actual spectrum of the matrix 
model. In section 5, we review the exact spectrum of the model for $\mu = 
\infty$ and describe the complete set of $SU(4|2)$ multiplets that it contains. 
In section 6, we apply our representation theory reasoning to determine which 
states have protected energies. We first identify all doubly atypical (and 
therefore exactly protected) multiplets in the spectrum. These include the 
vacuum states plus infinite towers of excited states above each vacuum. 
All of these states must be present 
with the same 
energy (in units of $\mu$), for all values of $\mu$. Furthermore, the doubly 
atypical spectrum about any given vacuum has a well defined large $N$ limit, so 
we conclude that these are exact quantum states of M-theory in the pp-wave 
background.

We then analyze the remaining atypical multiplets (which have the possibility of 
pairing up). We find some representations whose complementary multiplet is not 
present for $\mu = \infty$ and therefore cannot receive an energy 
shift as $\mu$ is varied from infinity. There are also pairs of multiplets in 
the $\mu = \infty$ that can combine to form typical multiplets. By explicit 
perturbative calculation we provide an example of one such pair which does 
receive an energy shift (and therefore must combine into a typical multiplet), 
as well as other such pairs of multiplets which do not receive an energy shift 
at leading order. As a check, we also verify a vanishing energy shift at leading 
order in perturbation theory for certain states that we predict are protected.
Finally, based on representation theory we argue that for the single 
membrane 
vacuum, the leading perturbative energy shift for all states (including those in 
typical multiplets) must display cancellations leaving a result that is finite 
in the large $N$ limit.

In section 7, we clarify the relation between atypical multiplets and BPS states 
(annihilated by one or more supersymmetry generators). We show that all BPS 
states lie in atypical multiplets and that all atypical multiplets contain BPS 
states. However, atypical multiplets generally contain some states which are not 
BPS. In fact, certain non-BPS atypical states carry no charges at all and yet 
have protected energies.

Finally, we offer some concluding remarks in section 8 and technical results in 
a few appendices.

\newpage

Note added: after this work was completed, the paper \cite{kp} appeared 
which partially overlaps with section 6 of this work. For other recent work on 
the pp-wave matrix model, see \cite{recent}.

\section{Symmetry algebra}

The symmetry algebra for the matrix model was discussed in \cite{dsv, bmn}. The 
bosonic generators are the light-cone translation generators $P^+$ (realized 
trivially as $P^+ = N/R$) and $P^-$ (the matrix model Hamiltonian), the $SO(3)$ 
and $SO(6)$ rotation generators $M^{ij}$ and $M^{ab}$, and the creation and 
annihilation operators $a_i, a_a$ associated with the center of mass harmonic 
oscillator. The fermionic generators include 16 simple generators $q$ which 
affect the overall polarization state, as well as the 16 non-trivial generators 
$Q$ which anticommute to give the Hamiltonian and rotation generators. 

The generators $P^+$, $a_i$, $a_a$ and $q$ act only on the $U(1)$ part of the 
theory which decouples, as discussed in \cite{dsv}. In this paper we will focus 
on the superalgebra generated by the remaining non-trivial generators $Q$, $H$, 
$M^{ij}$ and $M^{ab}$, with commutation relations 
\beas
\{ Q^{\dagger I \alpha}, Q_{J \beta} \} &=& 2 \delta^I_J \delta^\alpha_\beta H   
- {\mu \over 3} \epsilon^{ijk} \sigma^k_{\beta} {}^\alpha \delta^I_J
M^{ij} - {i \mu \over 6} \delta^\alpha_\beta ({\sf g}^{ab})_J {}^I M^{ab}\\
\left[H, Q_{I \alpha} \right] &=& {\mu \over 12} Q_{I \alpha}\\
\eeas
and additional commutators between $M$'s and $Q$'s appropriate to the fact that 
$Q_{I \alpha}$ transforms in the ${\bf (4,2)}$ of $SO(6) \times SO(3)$. 

This superalgebra satisfies all of the 
conditions\footnote{The conditions are that the algebra $G$ is simple, the 
bosonic subalgebra is reductive, and that there exists a non-degenerate 
invariant bilinear form on $G$.} for a ``basic classical Lie 
superalgebra,'' all of which have been 
classified by Kac \cite{kac}. As for the bosonic simple Lie algebras, these fall 
into several infinite series as well as a number of exceptional superalgebras. 
Among these superalgebras, there is precisely one whose bosonic subalgebra 
matches ours ($SO(6) \times SO(3) \times U(1)_H \sim SU(4) \times SU(2) \times 
U(1)_H$), namely the algebra $A_{3,1}$ whose compact form is known as 
$SU(4|2)$.\footnote{Actually, our Hamiltonian corresponds to a non-compact 
$U(1)$ generator, but this will make no difference to the representation 
theory.} 

For any values of the parameters $N$ and $\mu$, the spectrum of the matrix model 
must therefore lie in (finite-dimensional\footnote{We will demonstrate this 
below.}) representations of $SU(4|2)$. In particular, since the Hamiltonian is 
among the $SU(4|2)$ generators, the energy spectrum of states for given $N$ and 
$\mu$ is completely determined by which $SU(4|2)$ representations are present.
As a result, the energies of states in a given representation can only change as 
a function of $\mu$ if there are nearby representations with the same $SU(4) 
\times SU(2)$ state content but different energies.

In the next section, we will see that physically allowed representations of 
$SU(4|2)$ come in two types, known as typical and atypical (depicted
in Figure \ref{reps}). Typical 
representations lie along one-parameter families of representations which differ 
only by their energy eigenvalue. States in these representations can therefore 
shift up or down along the one parameter trajectories as a function of $\mu$ and 
therefore have energies which vary as a function of the parameters. On the other 
hand, atypical representations are isolated in the sense that there are no 
nearby representations with the same $SU(4) \times SU(2)$ state content. 
Physical states in these representations therefore have energies which are fixed 
as we vary $\mu$, except in special circumstances in which two such atypical 
representations combine to form a typical representation which may then shift to 
nearby typical representations with different energy.

By understanding the representations theory of $SU(4|2)$, and then determining 
precisely which representations are present at $\mu = \infty$ where the 
complete spectrum is known \cite{dsv}, we will be able to prove that certain 
infinite towers of states have energies that are protected as we vary $\mu$ away 
from $\mu = \infty$. This will provide precise information about the spectrum 
of the matrix model even for small $\mu$ where perturbation theory is 
inapplicable and also in the large $N$ limit defining M-theory on the pp-wave 
background.

We now turn to a discussion of the relevant properties of $SU(4|2)$ 
representations.

\section{Representations of $SU(4|2)$}

In this section, we review various aspects of the representation theory of 
$SU(4|2)$ that will be relevant to the matrix model. For a much more complete 
treatment, the reader is encouraged to refer to the article by Kac 
\cite{kac}, as well as further developments in \cite{bars,jakobsen}. 

The algebra $SU(4|2)$ naturally decomposes into subspaces with specific 
eigenvalues for the $U(1)$ generator (energy),
\[
{\cal G} = {\cal G}_0 \oplus {\cal G}_1 \oplus {\cal G}_{-1} 
\]
where ${\cal G}_0$ is the bosonic subalgebra $\{ M^{ab}, M^{ij}, H \}$ (whose 
generators all commute with 
H), and ${\cal G}_1$ and ${\cal G}_{-1}$ describe fermionic generators with 
positive and negative $H$ eigenvalues $\pm \mu/12$ ($Q_{I \alpha}$  and 
$Q^{\dagger I \alpha}$ respectively).   

Any given representation of $SU(4|2)$ splits up into a set of irreducible 
representations of $SU(4) \times SU(2)$ each labelled by an energy (the 
eigenvalue of the U(1) generator). Acting on a given state with fermionic 
generators in ${\cal G}_1$ or ${\cal G}_{-1}$ leads to states in 
different $SU(4) \times 
SU(2)$ representations with higher or lower energy. Since $\{ Q_{I \alpha}, Q_{J 
\beta} \} = 0$, our physical states will always lie in finite dimensional 
representations. Explicitly, if $|\psi_i \rangle$ are the states in a given 
$SU(4) \times SU(2)$ representation, the states 
\be
\label{basis}
|\psi_i ; \{ \epsilon_j, \tilde{\epsilon}_k \} \rangle \equiv 
(Q^\dagger_1)^{\epsilon_1} \cdots 
(Q^\dagger_8)^{\epsilon_8} (Q_1)^{\tilde{\epsilon}_1} \cdots 
(Q_8)^{\tilde{\epsilon}_8} | \psi_i \rangle \\
\ee
with $\epsilon_j, \tilde{\epsilon}_j = 0,1$ must be a complete basis of states 
for the full $SU(4|2)$ representation. To see this, note that using the 
commutation relations of the algebra, any product of $SU(4|2)$ generators acting 
on a state $|\psi_i \rangle$ may be rearranged to a sum of states of the form 
(\ref{basis}) by bringing all of the $Q^\dagger$'s to the left, then all of the 
$Q$s to the left of any remaining bosonic generators. Any remaining bosonic 
generators acting on $|\psi_i \rangle$ simply give a linear combination 
$c_{ij}|\psi_j \rangle$.

The representations appearing in the matrix model spectrum must not only be 
finite-dimensional, but all states must have positive energies and the 
representation must be unitarizable (i.e. admit a positive-definite inner 
product). Thus, we are interested in the finite-dimensional, positive-energy, 
unitarizable representations of $SU(4|2)$. We now proceed with a description of 
these representations.

\subsection{Highest weight characterization of representations}

\begin{figure}   
\vskip -0.3 cm
\centerline{\epsfysize=0.4truein \epsfbox{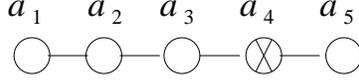}}
\vskip -0.3 cm
\caption{Kac-Dynkin diagram for $SU(4|2)$ }
\label{dynkin}
\end{figure}

The construction of representations for our superalgebra proceeds much like the 
familiar case of simple Lie algebras. We begin by choosing a maximal 
set of commuting bosonic generators from ${\cal G}_0$ which we denote by $H_i$. 
The 
remaining bosonic and fermionic generators may be chosen to be eigenvectors of 
the $H_i$ whose eigenvalues we call the roots. We may divide the roots into 
``positive'' and ``negative'' such that generators with positive roots together 
with the $H_i$ form a maximal subalgebra. Among the positive roots, there are 
five ``simple'' positive roots which cannot be written as sums of other positive 
roots. Three correspond to simple positive roots of the $SU(4)$ subgroup, one is 
a 
simple positive root for $SU(2)$,  and the final simple positive root 
corresponds 
to a fermionic generator. There exists a special choice of $H_i$ (the Dynkin 
basis) with the following property. For each simple positive root $\alpha_i$,  
we may choose generators $E_i$ and $F_i$ with roots $\alpha_i$ and $-\alpha_i$ 
such that 
\[
[H_i, H_j] = 0 \qquad [E_i, F_j] = H_i \delta_{ij} \qquad [H_i, E_j] = a_{ij} 
E_j \qquad [H_i, F_j] = -a_{ij} F_j
\]
and the Cartan Matrix $a_{ij}$ is given by
\[
a = \left( \ba{ccccc} 2 & -1 & & & \\ -1 & 2 & -1 & & \\ & -1 & 2 & -1 & \\ & & 
-1 & 0 & 1 \\ & & & -1 & 2 \ea \right)
\]
Associated to this Cartan matrix is a Kac-Dynkin diagram shown in
Figure \ref{dynkin}, 
where the fourth node corresponds to the fermionic simple positive root and in 
general the $i$th and $j$th nodes are joined by $|a_{ij} a_{ji}|$ lines. Note 
that the generators $H_1, H_2, H_3$ are a Cartan subalgebra of $SU(4)$ in the 
Dynkin basis while $H_5$ is a Cartan generator of $SU(2)$. Correspondingly, the 
left three nodes of the Kac-Dynkin diagram in Figure \ref{dynkin} give
the Dynkin diagram 
for $SU(4)$ while the right node is the Dynkin diagram for $SU(2)$.

In terms of this Dynkin basis, it is now straightforward to describe the finite 
dimensional representations. As usual, we may choose a basis of states in any 
representation such that the basis elements are eigenvectors of the Cartan 
subalgebra, and we call their eigenvalues the weights. For any finite 
dimensional 
representation, there exists a unique state with ``highest weight'' $\Lambda$ 
which is annihilated by all positive roots. We will denote this highest weight 
in the Dynkin basis by
\[
\Lambda = (a_1, a_2, a_3 |a_4| a_5) 
\]   
The finite dimensional representations are precisely those for which $a_1$, 
$a_2$, $a_3$, and $a_5$ are non-negative integers (while $a_4$ can be an 
arbitrary real number). Conversely, there exists a unique finite-dimensional 
irreducible representation corresponding to every such highest 
weight.\footnote{In general, for certain highest weights of this type, there may 
also exist representations which are not fully reducible, however these 
representations are not physically relevant since the realization of the 
superalgebra in the matrix model is in terms of hermitian generators.}

Acting on this highest weight state with the bosonic generators, we generate an 
irreducible representation $V_0(\Lambda)$ of $SU(4) \times SU(2)$ described by 
$SU(4)$ Dynkin labels $(a_1, a_2, a_3)$ and $SU(2)$ Dynkin label $a_5$ (spin 
$a_5/2$) with energy 
\be
\label{v0energy}
h \equiv {H \over \mu} = {1 \over 3}({1 \over 4} a_1 + {1 \over 2} a_2 + {3 
\over 4} a_3 + a_4 - {1 
\over 2} a_5)
\ee
The remaining states in the $SU(4|2)$ representation are now obtained by acting 
on 
these states with negative fermionic generators. These are exactly the 
generators $Q_{I \alpha}$ in ${\cal G}_{1}$, all of which have positive energy 
$h = 
1/12$, so the highest-weight representation $V_0$ is always the unique 
irreducible representation with lowest energy among the $SU(4) \times SU(2)$ 
representations in the full $SU(4|2)$ multiplet. Calling $|\psi_i \rangle$ 
the states in the highest weight representation, a basis of the full $SU(4|2)$ 
representation is then given by 
\[
|\psi_i ; \{ \epsilon_j \} \rangle \equiv (Q_1)^{\epsilon_1} \cdots 
(Q_8)^{\epsilon_8} | \psi_i \rangle
\]
where $\epsilon_j = 0,1$.\footnote{This follows immediately from the discussion 
above and the fact that $Q^{\dagger I \alpha} | \psi_i \rangle = 0$ for states 
in the lowest energy representation.}

If all these states are non-zero and independent, the representation is called 
``typical'', otherwise it is called ``atypical''. In terms of the highest 
weight, 
Kac showed that a representation is atypical if and only if 
\bea
a_4 &\in& \{ a_5 +1, a_5-a_3, a_5-a_3-a_2-1, a_5-a_3-a_2-a_1-2 \} \cr
&& \qquad \cup \{0, - a_3-1, -a_3-a_2-2, -a_3-a_2-a_1-3 \}
\label{atypical}
\eea
In the case that $a_4$ coincides with an element of each of the two sets on the 
right, the representation is known as ``doubly atypical'', and has additional 
special properties which we will describe later.

For typical representations, it is clear that we will have $SU(4) \times SU(2)$ 
representations at 9 equally spaced energy levels, and the dimension of the 
complete representation is 
\[
dim(V(\Lambda)) = 2^8 \cdot dim(V_0(\Lambda)) \; .
\]
Atypical representations always have fewer than 9 energy levels, and the 
dimension is less than the right-hand side of this formula. Explicit dimension 
formulae for atypical representations will be given in Figure \ref{dims}.

Among these finite dimensional irreducible representations, only a subset are 
compatible with unitarity and positive energy. These unitarizable 
representations have been characterized in \cite{jakobsen}. The conditions for 
unitarity may be given simply in terms of the highest weights as
\be
\ba{ll}
a_4 \in [a_5 + 1, \infty) &\qquad  a_5 \ge 1 \\
a_4 \in \{0\} \cup [1, \infty) &\qquad a_5 = 0 
\ea
\label{unitary}
\ee
In other words, among $SU(4|2)$ representations with a given lowest energy 
$SU(4) 
\times SU(2)$ irrep, any representation whose energy is greater than the 
highest energy atypical representation is unitarizable. In addition, for 
representations in which the lowest energy $SU(4) \times SU(2)$ irrep has 
trivial $SU(2)$ part ($a_5=0$), there is an additional atypical representation 
with lower energy that is unitarizable. These conditions are depicted in Figure 
\ref{reps}. 

To summarize, the physically allowed representations are those corresponding to 
highest weight $\Lambda = (a_1, a_2, a_3|a_4|a_5)$ such that $a_1$, $a_2$, 
$a_3$, and $a_5$ are non-negative integers and $a_4$ satisfies the conditions 
given in Eq.(\ref{unitary}). 

\subsection{Tensor representations and supertableaux}

In the case of ordinary Lie algebras, all finite dimensional representations may 
be obtained as tensor products of certain fundamental representations. This is 
not true for superalgebras (since the representation labels include a continuous 
parameter), however tensor representations will play a special role in our 
analysis of the Matrix model spectrum (all representations at $\mu = \infty$ are 
of this type), so we discuss them now. For more details, see the discussion by 
Bars et. al. \cite{bars}.

To describe the fundamental representation, we note that the superalgebra 
$SU(4|2)$ may be represented by matrix generators of the form
\[
{\cal H} = \left( \ba{cc} A & \theta \\ \theta^\dagger & B \ea \right)
\]
where $A$ and $B$ are hermitian $4 \times 4$ and $2 \times 2$ matrices with $\tr 
A = \tr B$ and $\theta$ is a $4 \times 2$ matrix of complex Grassman numbers 
(arbitrary linear combinations of the generators $Q_{I \alpha}$). Then the 
vector upon which this matrix acts defines the fundamental representation. 

We may denote such a vector by $\phi_A$ where the index $A$ takes values in $(I, 
\alpha)$ where $I$ is a fundamental index of $SU(4)$ and $\alpha$ is a 
fundamental 
index of $SU(2)$. 
\[
\phi = \left( \ba{c} \phi_I \\ \phi_\alpha \ea \right)
\]
It is clear that the states $\phi_I$ and $\phi_\alpha$ have opposite statistics 
since they are exchanged by the fermionic generators. In our discussion, we will 
take $\phi_I$ to be bosonic and $\phi_\alpha$ to be fermionic, but the other 
choice leads to an equivalent set of representations.

To determine the highest weight corresponding to this representation, we note 
that the matrix forms of the Cartan generators in the Dynkin basis described 
above are 
\beas
H_1 &=& {\rm diag}(1,-1,0,0,0,0)\\
H_2 &=& {\rm diag}(0,1,-1,0,0,0)\\
H_3 &=& {\rm diag}(0,0,1,-1,0,0)\\
H_4 &=& {\rm diag}(0,0,0,1,1,0)\\
H_5 &=& {\rm diag}(0,0,0,0,1,-1)\\
\eeas
so the $U(1)$ generator measuring energy is
\[
h = {\rm diag}({1 \over 12},{1 \over 12},{1 \over 12},{1 \over 12},{1 \over 
6},{1 \over 6})
\]
while the $SU(4)$ and $SU(2)$ subalgebras correspond to traceless generators in 
the upper left and lower right blocks respectively. Under the bosonic 
subalgebra, the fundamental representation thus splits into 
the ${\bf (4,1)} = (1,0,0) \times (0)$ representation of $SU(4) \times SU(2)$ 
with energy $h=1/12$ and the ${\bf (1,2)} = (0,0,0) \times (1)$ representation 
with energy $h=1/6$. The former representation, of lower energy, is the highest 
weight representation $V_0$, so we may immediately deduce that $(a_1, a_2, a_3) 
= (1,0,0)$ while $a_5 = 0$, and using the relation (\ref{v0energy}) with
$h=1/12$ 
we find $a_4 = 0$. Thus, the fundamental representation is 
\[
\Lambda = (1,0,0|0|0)
\]  
Higher tensor representations may be formed just as for $SU(N)$, by considering 
objects with multiple $A$ indices symmetrized in various ways.

At this point we should note that as for $SU(N)$, there is also an 
anti-fundamental representation which may be obtained as the complex conjugate 
of 
the fundamental representation and has highest weight $(0,0,0|0|1)$. Unlike the 
SU(N) case, the anti-fundamental representation here cannot be obtained from 
tensor products of the fundamental representation since there is no invariant 
tensor analogous to the epsilon tensor for $SU(N)$. Thus, to obtain the most 
general tensor representations for $SU(4|2)$, we must include both fundamental 
and 
anti-fundamental indices. However, it turns out that all tensor representations 
involving anti-fundamental indices either contain negative energy states or are 
not unitarizable,\footnote{This will follow from the discussion below. We should 
note, however, that while states in the matrix model must be in positive-energy 
unitary representations, physically interesting operators may be in more general 
representations. Indeed, the superalgebra itself is in the adjoint 
representation, a tensor representation corresponding to one fundamental and one 
anti-fundamental index.} so we 
will not consider them here. Thus, henceforth when we refer to tensor 
representations, we will mean tensor representations with only fundamental 
indices.

Since the various ways of symmetrizing the indices in a tensor representation 
are labelled by representations of the permutation group, we may label tensor 
representations of $SU(4|2)$ by Young tableaux, which we will call
supertableaux following \cite{bars}.
We use slashed boxes to distinguish them from ordinary tableaux 
which we will use to describe the $SU(4)$ and $SU(2)$ subgroups. For example, an 
object with two antisymmetrized super-indices $\phi_{[AB]}$ is denoted by the 
supertableau
\begin{figure}[ht]
\vskip -0.3 cm
\centerline{\epsfysize=0.25truein \epsfbox{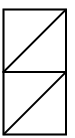}}
\vskip -0.3 cm
\label{2boxes}
\end{figure}
\newline 
The decomposition of a given tensor representation of $SU(4|2)$ into individual 
$SU(4) \times SU(2)$ representations corresponds to the possible ways of 
assigning the super-indices $A_i$ either to $SU(4)$ or $SU(2)$ fundamental 
indices. 
The energy of a given $SU(4) \times SU(2)$ irrep in the decomposition 
is given by 
\be
\label{energy}
h = {1 \over 12} n_4 + {1 \over 6 } n_2
\ee
where $n_4$ and $n_2$ are the number of $SU(4)$ and $SU(2)$ indices 
respectively.
For example, the tensor $\phi_{[AB]}$ decomposes into $SU(4) \times SU(2)$ 
tensors $\phi_{[IJ]}$ with energy $h = 1/6$, $\phi_{I \alpha}$ with energy 
$h=1/4$ and $\phi_{(\alpha \beta)}$ with energy $h=1/3$. Note that since 
the 
SU(2) indices are fermionic, the antisymmetrization of $AB$ becomes 
symmetrization of $\alpha \beta$. It is convenient to represent this 
decomposition in terms of tableaux as
\begin{figure}[ht]
\vskip -0.3 cm
\centerline{\epsfysize=0.3truein \epsfbox{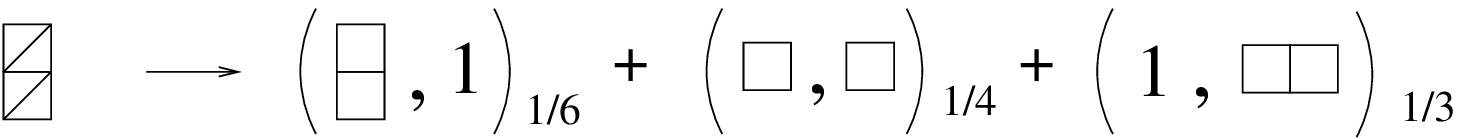}}
\vskip -0.3 cm
\label{2decomp}
\end{figure} 
\newline
The decomposition of more general $SU(4|2)$ representations into $SU(4) \times 
SU(2)$ representations may be efficiently carried out pictorially as explained 
in \cite{bars}.

\begin{figure}
\centerline{\epsfysize=0.9truein \epsfbox{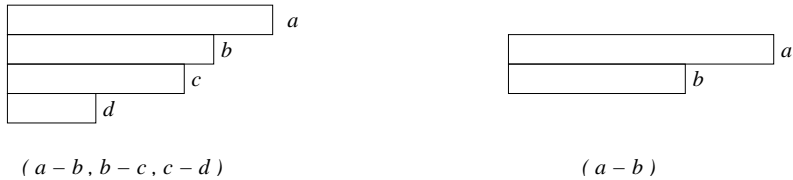}}
\caption{Allowed tableaux and highest weights for $SU(4)$ and $SU(2)$.}
\label{minusb}
\end{figure}

The restrictions on the allowed tableaux for $SU(4)$ and $SU(2)$ lead to 
restrictions on the allowed supertableaux for $SU(4|2)$. Recall that for 
$SU(N)$,  
we 
may have no more than $N$ antisymmetrized indices, since a fundamental index can 
take $N$ possible values. Thus, the maximum height of an $SU(N)$ tableau is $N$. 
Also, since $N$ antisymmetrized indices may be contracted with an invariant 
epsilon tensor to give a scalar, there is an equivalence between tableaux which 
allows one to eliminate any columns with $N$ boxes. The allowed tableaux for 
$SU(4)$ and $SU(2)$ are depicted in Figure \ref{minusb} along with the 
corresponding highest weights in the Dynkin basis. 

\begin{figure}
\centerline{\epsfysize=5truein \epsfbox{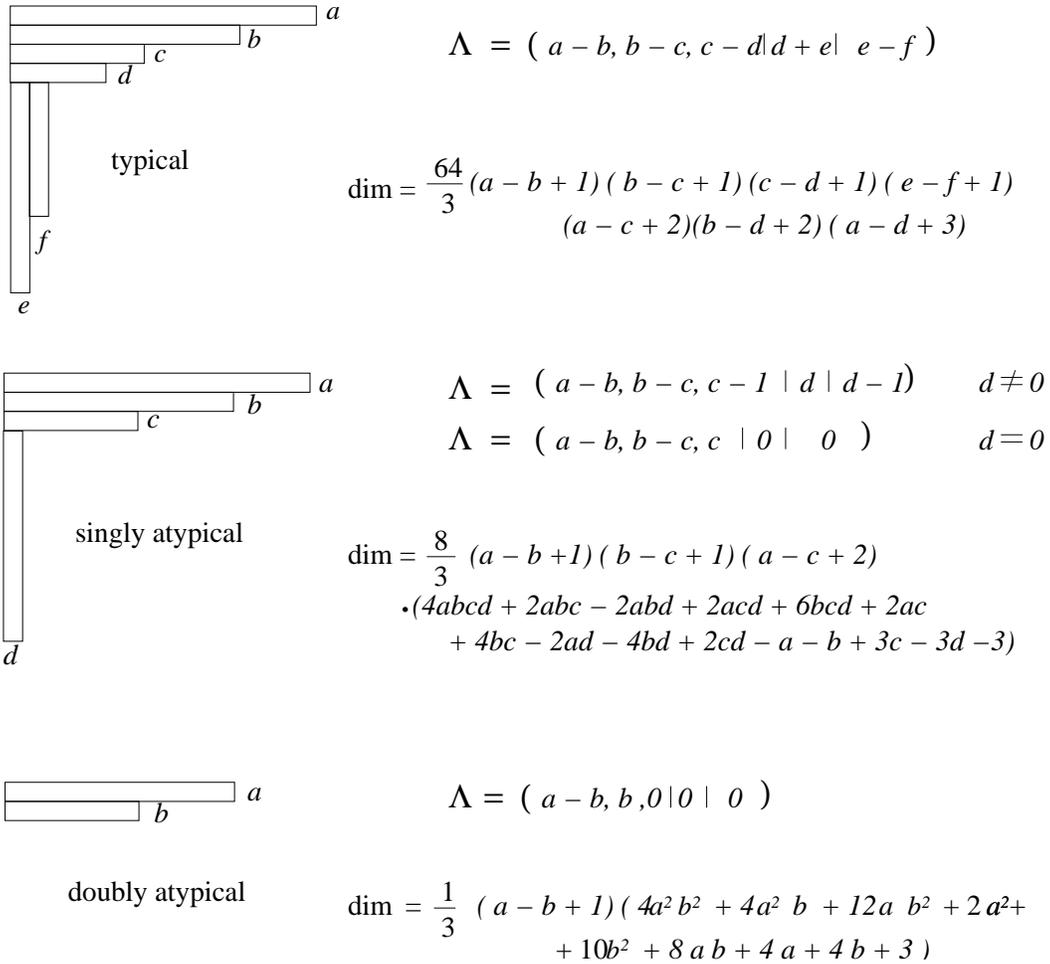}}
\caption{Supertableaux, highest weights, and dimensions for general
 tensor representations of $SU(4|2)$}
\label{dims}
\end{figure}

The condition on $SU(4|2)$ supertableaux is that there must exist at 
least one way to decompose the supertableau into allowed $SU(4)$ and $SU(2)$ 
tableaux. Pictorially, the condition is simply that the third column must have 
no 
more than four boxes. Otherwise, any decomposition would yield either an $SU(4)$ 
tableau with height greater than 4 or an $SU(2)$ tableau with height greater 
than 3. 
The most general allowed supertableau for $SU(4|2)$ is depicted at the top of 
Figure \ref{dims}.

\begin{figure}
\centerline{\epsfysize=1truein \epsfbox{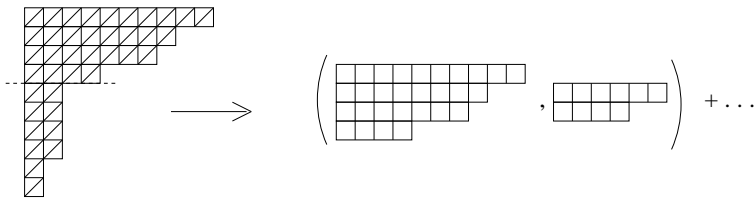}}
\caption{Lowest energy $SU(4) \times SU(2)$ multiplet in the
decomposition of a general $SU(4|2)$ multiplet.}
\label{gendecomp}
\end{figure}

It is straightforward to determine the highest weight corresponding to a given 
supertableau. Note that from Eq.(\ref{energy}), the lowest energy $SU(4) 
\times 
SU(2)$ representation for a given supertableau is obtained by assigning as few 
indices to $SU(2)$ as possible. This is accomplished by associating only those 
boxes which are below the fourth row to $SU(2)$,  with the complete set of 
boxes in the first four rows forming the $SU(4)$ tableau. Since the $SU(2)$ 
indices are fermionic, the $SU(2)$ tableau is obtained from the part of the 
supertableau below the fourth row by a flip on its diagonal, thus 
exchanging symmetrization 
and antisymmetrization. This decomposition is depicted in Figure 
\ref{gendecomp}. 

Given this 
lowest energy $SU(4) \times SU(2)$ representation $V_0$, the highest weight 
components $(a_1,a_2,a_3)$ and $(a_5)$ are the Dynkin labels for the $SU(4)$ and 
SU(2) representations in $V_0$. To determine $a_4$, we may use the formula 
(\ref{v0energy}) where the energy $h$ is given by Eq.(\ref{energy}). In 
this 
case, $n_4$ 
and $n_2$ are the total number of
boxes in the supertableau in the
first four rows and in the remaining
rows respectively.

The general allowed supertableau and the corresponding highest weight are 
depicted at the top of Figure \ref{dims}. 
We note that there is an 
equivalence between 
supertableaux given by adding $k$ columns with four boxes and subtracting $k$ 
rows with two boxes, or in the notation of Figure \ref{dims},
\be
\label{equ}
(a,b,c,d,e,f) \rightarrow (a+k,b+k,c+k,d+k,e-k,f-k)
\ee
where $k$ is any integer such that the resulting supertableau is sensible. By 
this equivalence, any tensor representation may be denoted uniquely using a 
tableau with no more than one box in the fifth row.

\vskip .3cm

Comparing the highest weight with the conditions for atypicality, we find that 
the atypical tensor representations are exactly those corresponding to tableau 
whose second column has no more than three boxes, while the the doubly atypical 
representations correspond to those whose first column has no more than two 
boxes. The general supertableaux for singly and doubly atypical representations 
are also depicted in Figure \ref{dims} along with their highest weights. 
Also, using the methods in \cite{bars}, we have computed the dimensions for each 
type of tensor representation and included these formulae in Figure \ref{dims}. 
From \cite{bars}, one may also compute the relative number of bosons and 
fermions in these representations, and one finds that that they are equal for 
typical and singly atypical representations, but different for doubly atypical 
representations, with an excess of $a-b+1$ bosonic states, where $a$ and 
$b$ are 
the number of boxes in the first and second rows of the doubly atypical tableau.

\begin{figure}
\centerline{\epsfysize=4truein \epsfbox{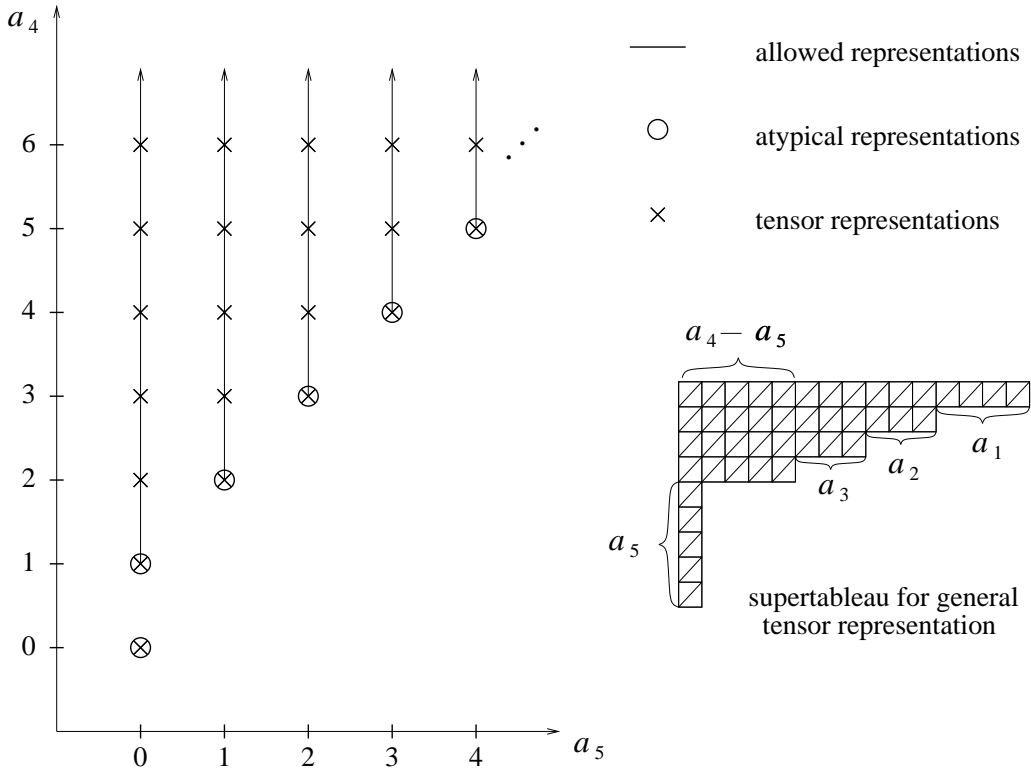}}
\caption{ Physically allowed representations of $SU(4|2)$ for a given
$(a_1,a_2,a_3)$}
\label{reps}
\end{figure}

We may now compare the highest weights in Figure \ref{dims} with the 
conditions for 
positive energy and unitarity in Eq.(\ref{unitary}). We find that every 
tensor 
representation is positive energy and unitarizable, and further, that every 
positive-energy unitarizable finite-dimensional irreducible representation with 
integer highest weight is a tensor representation.\footnote{Since general 
tensor 
representations involving both fundamental and anti-fundamental indices also 
correspond to integer highest weights, they must have negative energy states or 
be non-unitary, as asserted above, since all of the positive-energy unitarizable 
integer-weight representations are tensor representations involving only 
fundamental indices.} This is summarized in Figure \ref{reps} which 
displays the 
complete set of physically allowable representations along with the discrete 
subset corresponding to tensor representations and the further subset 
corresponding to atypical representations.    

Finally, we note that tensor products between tensor representations may be 
computed from the supertableaux with the usual Littlewood-Richardson rules for 
computing tensor products of $SU(N)$ representations. In this case, when 
multiplying two tableaux, we keep only the resulting tableaux which are allowed 
tableaux of $SU(4|2)$. For example,
\begin{figure}[ht]
\vskip -0.3 cm
\centerline{\epsfysize=0.3truein \epsfbox{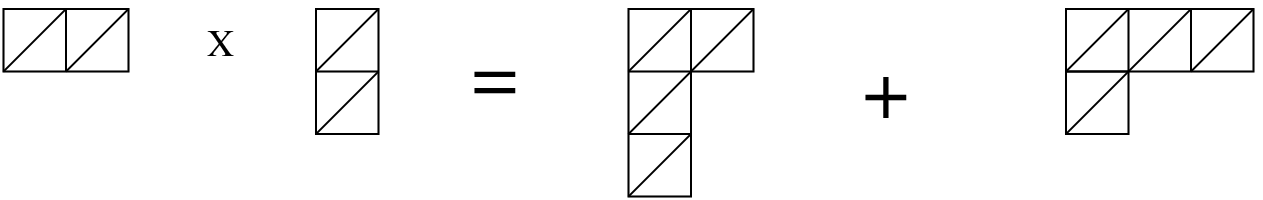}}
\vskip -0.3 cm
\end{figure} 

\section{Physical implications}

We have shown that for the matrix model describing M-theory on the
maximally supersymmetric pp-wave, the symmetry
algebra generated by the 16 non-trivial
supercharges $Q$, the $SO(3)$ and $SO(6)$ rotation generators $M$, and the
Hamiltonian $H$ is the basic classical Lie superalgebra $SU(4|2)$. In
the previous section, we reviewed the physically allowable
representations of $SU(4|2)$ and showed that these consist of one-parameter 
families of typical representations as well as a discrete set of atypical 
representations, as depicted in Figure \ref{reps}. The typical 
representations in a 
one-parameter family (corresponding to the vertical lines in the
Figure \ref{reps} 
excluding the lower endpoints) have identical $SU(4) \times SU(2)$ content and 
differ only by the overall energies of the states (the energy differences in a 
representation do not change). On the other hand, atypical representations have 
no nearby representations with the same $SU(4) \times SU(2)$ content but 
different energy.
 
For any fixed values of the parameters $N$ and $\mu$, the spectrum of the matrix 
model will consist of some discrete set of these $SU(4|2)$ representations which 
may include typical and atypical representations. We can denote this information 
by a set
\[
S  = \{ \Lambda_i \}(N,\mu)
\]
of the highest weights corresponding to the various representations. We would 
now like to understand what can happen to this spectrum of representations as we 
vary the parameter $\mu$. 

In general, the $SU(4) \times SU(2)$ quantum numbers of any given state can't 
change, and the energy may only vary continuously. For states in any typical 
representation, the energy may shift up or down (with the same shifts for all 
states in the representation) as we vary $\mu$ since there are nearby 
representations with the same $SU(4) \times SU(2)$ content with both higher and 
lower energy. This corresponds to continuously changing some of the highest 
weights in our set $S$ along the one-parameter families of Figure 
\ref{reps}.
On the other hand, states in atypical representations generally cannot receive 
any shift in energy since, there are no representations nearby in energy with 
the 
same $SU(4) \times SU(2)$ state content. However, it is possible that two (or 
more) atypical representations could combine into a typical representation 
or vice versa. This would correspond to a 
discontinuous transition in $S$ in which certain atypical highest weights appear 
or disappear. We will now determine for which representations this is possible.

Let $\mu = \mu_0$ be a point for which such a discontinuous transition occurs. 
That 
is, we assume that the states in a set of multiplets at $\mu$ arbitrarily close 
to 
$\mu_0$ arrange themselves into a different set of multiplets when $\mu$ 
reaches $\mu_0$.
Among the states involved in such a discontinuous transition, there will 
be some (or possibly more than one) state $|\phi \rangle$ of highest weight. The 
weight of this state (which must change continuously as we vary $\mu$) must be 
present as a highest weight in the set $S$ both at $\mu_0$ and away from it, but 
by assumption, the $SU(4) \times SU(2)$ content of the $SU(4|2)$ 
representation in which 
$|\phi \rangle$ sits will change as $\mu$ changes from $\mu_0$. From Figure 
\ref{reps}, we see that the only case in which the $SU(4) \times SU(2)$ 
content of a 
representation changes with an infinitesimal change in the highest weight is the 
situation in which a highest weight corresponding to a typical representation 
reaches the bottom of its one parameter family. At this point, the corresponding 
representation becomes atypical and will have less states than the nearby 
typical representation from which it arose. For the transition to be possible, 
the remaining states which drop out of this representation must form a separate 
representation (or representations) of $SU(4|2)$ with some lower highest 
weight(s). 

To see if this is possible, we consider a general one parameter family of 
typical representations corresponding to highest weight $\Lambda = (a_1, a_2, 
a_3| a_5 + 1 + \epsilon| a_5)$ where $\epsilon > 0$. These representations all 
have the same $SU(4) \times SU(2)$ content with energies depending linearly on 
$\epsilon$. Let $A = \{(R_i, h_i) \}$ be the $\epsilon \to 0$ limit of this 
set of $SU(4) \times SU(2)$ representations and energies. Similarly, we will 
have some set B of $SU(4) \times SU(2)$ representations and energies in the 
atypical representation with highest weight $(a_1, a_2, a_3| a_5 + 1| a_5)$. 
The set $B$ will be some subset of the set $A$, and we may define a set $C$ to 
be the elements of $A$ not in $B$. We then check whether the elements in $C$ 
match with some set of complete $SU(4|2)$ representations.

\begin{figure}
\vskip -0.3 cm
\centerline{\epsfysize=1truein \epsfbox{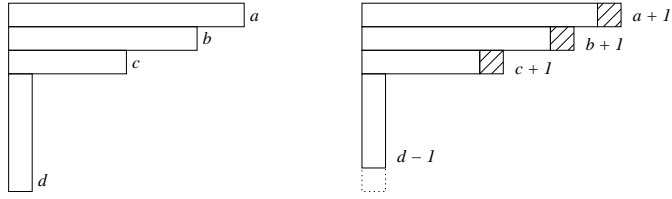}}
\vskip -0.3 cm
\caption{Atypical representations which can combine into a typical 
representation}
\label{combine}
\end{figure} 

It turns out that in every case, the elements 
of $C$ precisely correspond to the decomposition of a single atypical 
representation (whose highest weight may be found by finding the lowest energy 
representation in $C$). For $a_5 > 0$, the states of $C$ match exactly 
with the 
representation of highest 
weight $\Lambda = (a_1, a_2, a_3+1 | a_5 |a_5 - 1)$, while for $a_5 = 0$, 
the states of $C$ match exactly with the representation of highest weight 
$\Lambda = (a_1, a_2 , a_3 + 2 | 0 | 0 )$. Using the dimension
formulae in Figure \ref{dims}  it is straightforward to verify that
$dim(B)+dim(C)=dim(A)$ for the representations given. 

We may conclude that the complete set of possible discontinuous transitions is 
\be
(a_1, a_2, a_3| a_5 + 1 + \epsilon| a_5) \longleftrightarrow (a_1, a_2, 
a_3| a_5 + 1 | a_5) \oplus (a_1, a_2, a_3+1 | a_5 |a_5 - 1) 
\ee
for $a_5 > 0$ and 
\be
(a_1, a_2, a_3| 1 + \epsilon| 0) \longleftrightarrow (a_1, a_2, a_3| 1 | 
0) 
\oplus (a_1, a_2, a_3 + 2 | 0 | 0 )\ .
\ee
with $a_5 = 0$. In terms of the supertableaux, the pairs of atypical
representations which can 
combine to form a typical representation are those for which one of the 
tableaux 
has one less box in the first column and one more box in each of the first three 
rows, as depicted in Figure \ref{combine}.  

Thus, if a certain atypical representation exists in the spectrum at
some value of $\mu= \mu_0$, 
it may only receive an energy shift for nearby values of $\mu$ if a
complementary typical representation, obtained by the operation in
Figure \ref{combine} or its inverse, is also present in the spectrum
at $\mu = \mu_0$, and any such energy shift must be positive. 

\subsection{Exactly protected representations}

\begin{figure}
\vskip -0.3 cm
\centerline{\epsfysize=0.4truein \epsfbox{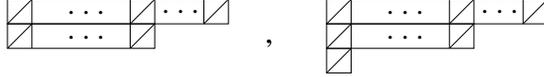}}
\vskip -0.3 cm
\caption{Atypical representations which can never combine}
\label{nosplit}
\end{figure}

There are certain representations, depicted in Figure \ref{nosplit},
which are not members of any pair of representations which can combine.
In particular, these include all doubly atypical representations 
(i.e. those corresponding to 
tableaux with less than three rows).\footnote{Actually, it is obvious
that doubly atypical representations cannot combine with other
representations to form typical representations since they all have
more bosons than fermions while all singly atypical and typical
representations have the same number of bosons and fermions.} 
We may conclude that if the spectrum 
contains any of these representations for some value of $\mu$, then they must be 
present for all other values of $\mu$, and the energies of states in these 
representations are protected from any perturbative or non-perturbative shifts.

\subsection{A family of supersymmetric index theorems}

We will now show that in addition to the occupation numbers for each type of
multiplet in Figure \ref{nosplit}, there are additional exactly protected
quantities involving the atypical multiplets not in Figure \ref{nosplit}.

\begin{figure}
\vskip -0.3 cm
\centerline{\epsfysize=0.9truein \epsfbox{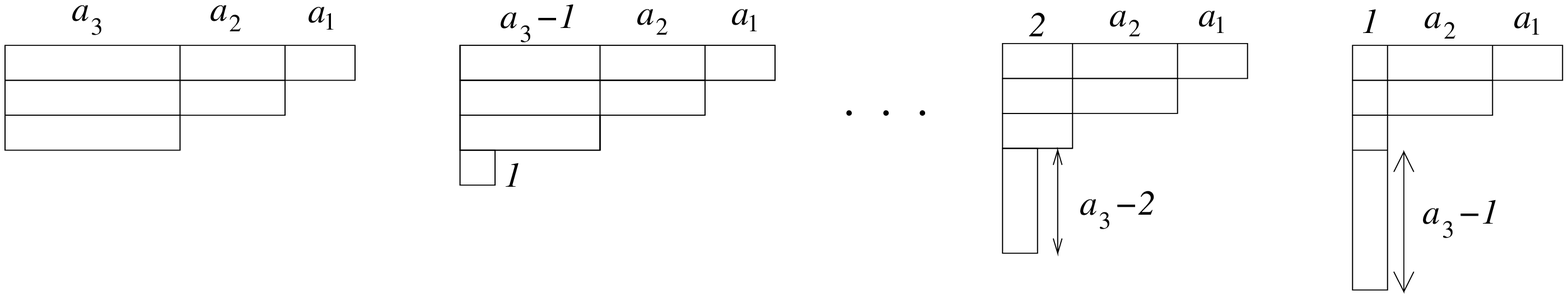}}
\vskip -0.3 cm
\caption{Chains of atypical representations in which nearest neighbors
may combine}
\label{chains}
\end{figure}

We first note that all atypical multiplets may be arranged into finite
chains as depicted in Figure \ref{chains}, where moving to the left in
the chain corresponds to performing the operation in Figure
\ref{combine}. The leftmost multiplet in each chain corresponds to a
tableau with less than four rows, while the rightmost multiplet in every
chain corresponds to a tableau with less than two boxes in the third
row. Each atypical multiplet will appear in exactly one chain. The
chains may be labelled by the elements $(a_1,a_2,a_3)$ of the highest
weight for the leftmost multiplet (i.e. the highest weight of its lowest
energy $SU(4)$ representation), and the length of each chain will be
the greater of $a_3$ and 1.

Now, consider some chain of atypical multiplets $(A_1, \dots,
A_n)$. By definition, each pair $(A_i, A_{i+1})$ may combine and shift
along some one parameter family of typical multiplets which we denote
by $T_i$. From the discussion above, if $A_i$ is present in the spectrum at
$\mu = \mu_0$, it will be protected as $\mu$ is varied away from $\mu_0$
as long as neither of its nearest neighbor multiplets $A_{i-1}$  or
$A_{i+1}$ in the chain appear in the spectrum for $\mu =
\mu_0$. However, this multiplet is not
necessarily protected for all values of $\mu$. The reason is that one
of its neighbor representations may appear in the spectrum at some
$\mu = \mu_1$ without introducing a new $A_i$ multiplet if a
representation in $T_{i+1}$ or $T_{i-2}$ shifts to the bottom of its
one-parameter family and splits. For the $T_{i+1}$ case, we would then
have representations $A_i$, $A_{i+1}$, and $A_{i+2}$ all present at
$\mu = \mu_1$. As $\mu$ is varied further, it is then possible
for $A_i$ and $A_{i+1}$ to combine and shift to a representation in
$T_i$, leaving only $A_{i+2}$. Thus, the fact that a multiplet is
protected in the neighborhood of some point $\mu_0$ does not imply
that it is protected for all $\mu$. 

While the number ${\cal N}_i$ of multiplets $A_i$ in the spectrum can change as 
we vary $\mu$, the set $\{ {\cal N}_i \}$ cannot change arbitrarily. Taking into 
account all possible ways in which the multiplets $A_i$/$T_i$ can combine/split 
as we vary $\mu$, it is apparent that the most general possible change in the 
occupation numbers between two values of $\mu$ is a transformation
\[
({\cal N}_1, {\cal N}_2, \dots, {\cal N}_{n-1}, {\cal N}_n) \rightarrow ({\cal 
N}_1 + n_1, {\cal N}_2 + n_1 + n_2, \dots, {\cal N}_{n-1} + n_{n-2} + n_{n-1}, 
{\cal N}_n + n_{n-1})
\]
where $n_i$ are integers such that all values on the right-hand side are 
non-negative. There is precisely one linear combination of the occupation 
numbers 
invariant under all such transformations, namely the alternating sum
\[
{\cal I} = \sum_k (-1)^k {\cal N}_k \; .
\]
Therefore, this quantity defines a ``supersymmetric index'' 
which is exactly protected for all values of $\mu$.

In summary, given any chain of representations labelled by the $SU(4)$ highest 
weight $(a_1, a_2, a_3)$  of its leftmost multiplet, there exists an exactly 
protected quantity ${\cal I}(a_1,a_2,a_3)$ given by the alternating sum of 
occupations numbers of multiplets in the chain. Explicitly, we have
\bea
{\cal I}(a_1,a_2,a_3) = {\cal N}(a_1, a_2, a_3|0|0) &-& \sum_{n \ge 0} {\cal N}
(a_1,a_2,a_3-(2n+2)|2n+1|2n) \label{index}\\
&+& \sum_{n \ge 0} {\cal N}(a_1,a_2,a_3-(2n+3)|2n+2|2n+1) \nonumber
\eea
where ${\cal N}(\Lambda)$ denotes the number of multiplets in the spectrum with 
highest weight $\Lambda$.

We would now like to apply what we have learned to the actual physical spectrum 
of the matrix model, which we determined for $\mu = \infty$ in \cite{dsv}. By 
identifying which representations are present for $\mu = \infty$ we will be able 
to use the results of this section to see which states are protected
as $\mu$ moves away from $\infty$.

\section{$SU(4|2)$ representations in the matrix model spectrum at $\mu = 
\infty$}

In \cite{dsv}, the exact spectrum of the matrix model was determined in the $\mu 
= \infty$ limit. From this explicit construction, we will now determine 
which $SU(4|2)$ representations are present in this limit. In section
6, we will then use the results of section 4 to investigate which of
these multiplets have protected energies as we move away from $\mu = \infty$.

To begin, we briefly recall the construction in \cite{dsv}. For a given $N$ and 
general non-zero $\mu$, the matrix model contains a collection of isolated vacua 
corresponding to the various ways of dividing up the total DLCQ momentum $N$ 
into some number of distinct gravitons, which in this background appear as 
concentric giant graviton spheres. The radius of each fuzzy sphere is 
proportional to the number of units of momentum it carries. We may expand 
the matrix 
model action about any of these vacuum states, and for large $\mu$, we find a 
quadratic Hamiltonian with interaction terms suppressed by powers of $1 / \mu$. 
Thus, in the limit of large $\mu$, the various vacua become superselection 
sectors each described by a quadratic Hamiltonian which may be diagonalized 
explicitly to yield towers of oscillators that generate the spectrum. The 
spectrum of oscillators for the single-membrane vacuum and for a general vacuum 
are reproduced here in Tables 1 and 2. 

To determine which representations will be present in the spectrum, we first 
note that like the Hamiltonian, the supercharges $Q$ expanded about a given 
vacuum become quadratic in the large $\mu$ limit and are comprised of terms 
containing one creation operator and one annihilation operator. For
the single membrane vacuum, we find\footnote{The expression for a
general vacuum is identical except that the oscillators carry indices
$k,l$ corresponding to which block they arise from, and the spin sums
are those given in Table 2.} (more details are given in the Appendix A)
\bea\label{QQ}
Q_{I\alpha}&= &\sum_{j=1}^{N-1}
i\sqrt{{2\mu\over 3}} a_{j-1\ m}
\left(\matrix{
-\sqrt{j-m}\ {(\chi^{\dagger})_I}^{j-{1\over 2}\  m-{1\over 2}}  \cr
-\sqrt{j+m}\ {(\chi^{\dagger})_I}^{j-{1\over 2}\  m+{1\over 2}}  
}\right)_\alpha \cr \; \cr
&+ & \sum_{j=1}^{N-1} i{\sqrt{{\mu\over 3}}} {\sf g}^a_{IJ} a^{a\ 
\dagger}_{jm}
\left(\matrix{
-\sqrt{j+m}\ (\chi^J)_{j-{1\over 2}\ -m-{1\over 2}}  \cr
-\sqrt{j-m}\ (\chi^J)_{j-{1\over 2}\ -m+{1\over 2}}  }\right)_\alpha \cr \; \cr
&+ &\sum_{j=0}^{N-1} i{\sqrt{{\mu\over 3}}} {\sf g}^a_{IJ} a^{a}_{jm}
\left(\matrix{
\sqrt{j-m+1}\ \eta^{\dagger\ J}_{j+{1\over 2}\ m-{1\over 2}}  \cr
-\sqrt{j+m+1}\ \eta^{\dagger\ J}_{j+{1\over 2}\ m+{1\over 2}}  
}\right)_\alpha \cr \; \cr
&+ &\sum_{j=0}^{N-1} i{\sqrt{{2\mu\over 3}}} 
b^{\dagger}_{j+1\ m}\left(\matrix{
\sqrt{j+m+1}\ \eta_I^{j+{1\over 2}\ -m-{1\over 2}}  \cr
-\sqrt{j-m+1}\ \eta_I^{j+{1\over 2}\ -m+{1\over 2}}  
}\right)_\alpha \ .
\eea
Thus, under the 
action of the superalgebra on a given eigenstate in the $\mu = \infty$ limit, 
the total number of oscillators is preserved. As a result, the subsector of 
states 
with any given number of oscillators (which we will sometimes loosely
refer to as ``particles'') must arrange into some set of complete 
$SU(4|2)$ representations.  

To determine the representations in the spectrum, we will first determine the 
representations corresponding to the individual creation operators upon which 
the spectrum is built. The representations making up the rest of the
spectrum will be obtained by tensor products of these single 
oscillator representations.

\subsection{Single membrane vacuum}

We begin by determining the single-oscillator representations for the 
single-membrane vacuum using Table 1. 

We will use the fact that physically allowed 
representations of $SU(4|2)$ are completely determined by the energy and 
$SU(4) 
\times SU(2)$ representation of their lowest energy component. The lowest energy 
state in Table 1 is the U(1) oscillator $x_{00}$, whose $SU(4) \times SU(2)$ 
representation and energy are given by 
\begin{figure}[ht]
\vskip -0.3 cm
\centerline{\epsfysize=0.25truein \epsfbox{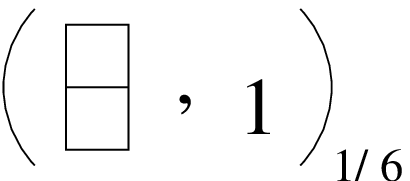}}
\vskip -0.3 cm
\label{onesixth}
\end{figure} 
\newline
it is easy to check that this matches with the 
lowest energy 
component of the $SU(4|2)$ representation described by the tableau 
\begin{figure}[ht]
\vskip -0.3 cm
\centerline{\epsfysize=0.25truein \epsfbox{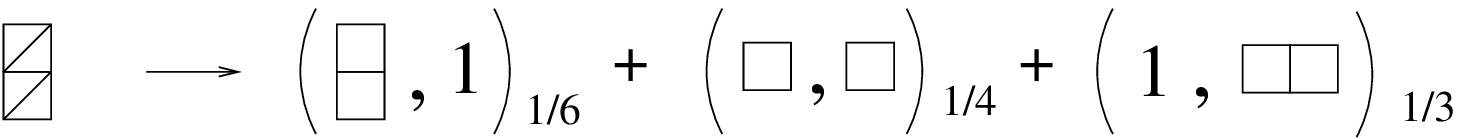}}
\vskip -0.3 cm
\label{}
\end{figure} 
\newline
We see that the remaining states in this representation match exactly 
with the 
single particle states corresponding to $\eta_{1 \over 2}$ and $\beta_1$, which 
are the remaining U(1) oscillators, i.e. all these modes are proportional 
to the identity matrix. Thus, the U(1) oscillators form a single 
representation with highest weight $(0,1,0|0|0)$. 

Apart from these states, the next lowest energy single-oscillator state is the 
energy $h=1/3$,  $SU(4) \times SU(2)$ singlet state corresponding to 
$\alpha_{00}$. 
From Figure \ref{dims}, we find that the $SU(4|2)$ representation 
corresponding to this $V_0$ is
\begin{figure}[ht]
\vskip -0.3 cm
\centerline{\epsfysize=0.4truein \epsfbox{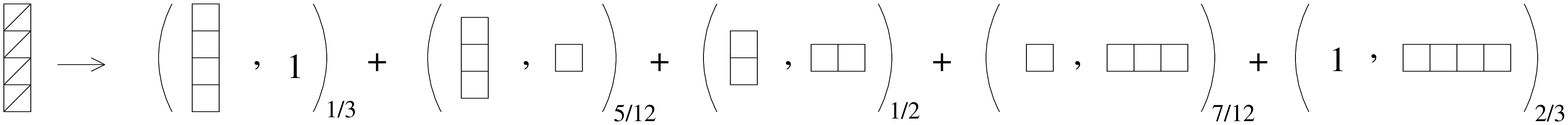}}
\vskip -0.3 cm
\end{figure} 
\newline
with highest weight $(0,0,0|1|0)$.
From Table 1, we find that the remaining four components of this 
representation 
must correspond to the oscillators $\chi_{1 \over 2}$, $x_1$, $\eta_{3 \over 
2}$, and $\beta_2$, which are precisely the oscillators in the $SU(2)$ theory.

\begin{figure}
\vskip -0.3 cm
\centerline{\epsfysize=1.9truein \epsfbox{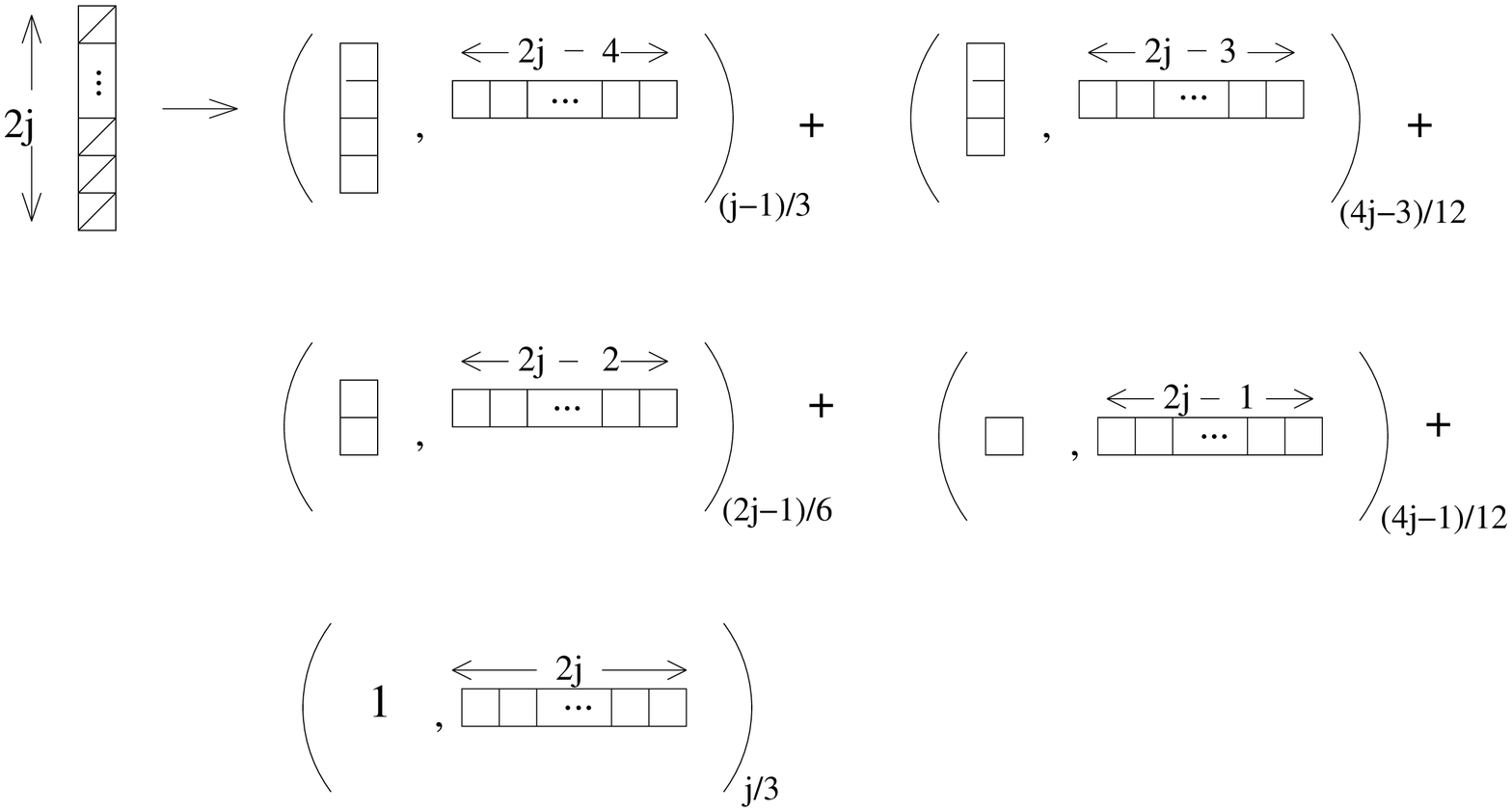}}
\caption{Decomposition of the general single oscillator multiplet}
\label{2jdecomp}
\vskip -0.3 cm
\end{figure}

Continuing in this way, we find that the remaining oscillators arrange into 
$SU(4|2)$ representations with highest weight $(0,0,0|2j+1|2j)$,
depicted in Figure \ref{2jdecomp}, where the individual $SU(4) \times
SU(2)$ representations in the figure correspond to 
$\alpha_j$, $\chi_{j+{1 \over 2}}$, $x_{j+1}$, $\eta_{j + {3 \over 2}}$, and 
$\beta_{j+2}$ respectively.\footnote{These are exactly the oscillators
which couple to matrix spherical harmonics $Y_{j+1 \; m}$ of spin $j+1$.}

\begin{figure}
\vskip -0.3 cm
\centerline{\epsfysize=0.7truein \epsfbox{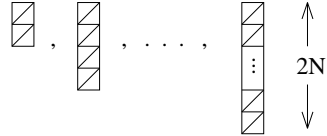}}
\vskip -0.3 cm
\caption{Supertableaux for all single oscillator multiplets for the single 
membrane vacuum}
\label{2Nboxes}
\end{figure}

Thus, in the matrix model with $P^+ = N/R$, the creation operators for the 
single-membrane vacuum make up the $N$ $SU(4|2)$ tensor representations given 
by the supertableaux of Figure \ref{2Nboxes}. 

Using the super-index notation, we may thus combine all of the oscillators 
from 
Table 1 into a set of ``super-oscillators''
\[
a^\dagger_{[A_1 A_2]}, a^\dagger_{[A_1 A_2 A_3 A_4]}, \dots  a^\dagger_{[A_1 
\cdots A_{2N}]} \; .
\]

\begin{table}
\begin{center}
\begin{tabular}{|c||c|c|c|c|c|}
\hline
{\rm Type} & {\rm Label} & {\rm Mass} & {\rm Spins} & $SO(6) \times 
SO(3)$ {\rm Rep} & {\rm Degeneracy} 
\\ \hline\hline
$S0(6)$ & $x^a_{jm}$ & ${1 \over 6} + {j \over 3}$ & $0 \le j \le N-1$ 
& $(6, 2j+1)$ & $6(2j+1)$ 
\\ \hline
$S0(3)$ & $\alpha_{jm}$ & ${1 \over 3} + {j \over 3}$ & $0 \le j \le N-2$ 
& $(1, 2j+1)$ & $(2j+1)$
 \\ 
  & $\beta_{jm}$ & ${j \over 3}$ & $1\le j \le N$ & $(1, 2j+1)$ & $(2j+1)$ 
\\ \hline
{\rm Fermions} & $\chi^I_{jm}$ & ${1 \over 4} + {j \over 3}$ & 
${1 \over 2} \le j \le N - {3 \over 2}$ & $(\bar{4}, 2j+1)$ & $4(2j+1)$ 
\\ 
   & $\eta_{I \; jm}$ & ${1 \over 12} + {j \over 3}$ & ${1 \over 2} \le j \le 
N - {1 \over 2}$ & $(4, 2j+1)$ & $4(2j+1)$ \\
\hline
\end{tabular} \caption{Oscillators for the single membrane vacuum}
\end{center}
\end{table}

\begin{table}
\begin{center}
\begin{tabular}{|c||c|c|c|c|c|}
\hline
{\rm Type} & {\rm Label} & {\rm Mass} & {\rm Spins} & $SO(6) \times SO(3)$
\\ \hline\hline 
$S0(6)$ & $(x_{kl}^a)_{jm}$ & ${1 \over 6} + {j \over 3}$ & 
${1 \over 2}|N_k - N_l| \le j \le {1 \over 2} (N_k + N_l) - 1$ & $(6,2j+1)$ 
\\ \hline 
$S0(3)$ & $\alpha_{kl}^{jm}$ & ${1 \over 3} + {j \over 3}$ & ${1 \over 
2}|N_k - N_l| - 1 \le j \le {1 \over 2} (N_k + N_l) - 2$ & $(1,2j+1)$ 
\\ 
 & $\beta_{kl}^{jm}$ & ${j \over 3}$ & ${1 \over 2}|N_k - N_l| +1 \le j 
\le {1 \over 2} (N_k + N_l)$ & $(1,2j+1)$  
\\ \hline
{\rm Fermions} & $\chi^{I \;jm}_{kl}$ & ${1 \over 4} + {j \over 3}$ & 
${1 \over 2}|N_k - N_l| - {1 \over 2} \le j \le {1 \over 2} (N_k + N_l) - 
{3 \over 2}$ 
& $(\bar{4},2j+1)$ \\  
  & $\eta_{I \; kl}^{jm}$ & ${1 \over 12} + {j \over 3}$ & ${1 \over 2}|N_k - 
N_l| 
+ {1 \over 2} \le j \le {1 \over 2} (N_k + N_l) - {1 \over 2}$ & $(4,2j+1)$ 
\\
\hline
\end{tabular} \caption{Oscillators for general vacua}
\end{center}
\end{table}

\begin{figure}
\vskip -0.3 cm
\centerline{\epsfysize=0.6truein \epsfbox{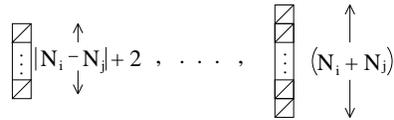}}
\caption{Supertableaux for single oscillator multiplets in a general vacuum}
\vskip -0.3 cm
\label{Reduce}
\end{figure}

\clearpage

\begin{figure}[ht]
\vskip -0.3 cm
\centerline{\epsfysize=0.75truein \epsfbox{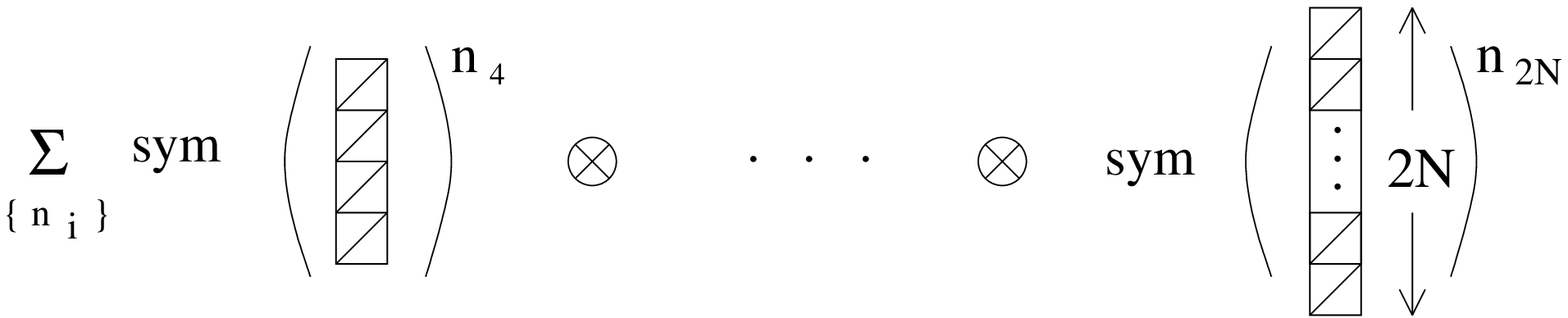}}
\vskip -0.3 cm
\caption{All $SU(4|2)$ multiplets in the single membrane vacuum for $SU(N)$. }
\label{nfour}
\end{figure}

The $SU(4|2)$ representations corresponding to states with multiple oscillators 
are those obtained by acting with arbitrary combinations of these 
super-oscillators on the Fock-space vacuum. The complete set of representations for 
the single membrane vacuum in the $SU(N)$ theory is therefore given by the 
tensor product in Figure \ref{nfour}.
Here, we have indicated that with multiple copies of similar oscillators, we 
must keep only representations in the symmetrized tensor product of these 
oscillators. Such tensor products may be computed easily for example using the 
group theory calculation software LiE \cite{lie}.

In the tensor product of Figure \ref{nfour}, we have not included a
factor corresponding to the 
$U(1)$ oscillators. In our analysis of which states are protected, we may work 
directly with the representations in the $SU(N)$ part of the theory, since the 
representations corresponding to the free $U(1)$ part of the theory cannot 
change as we vary $\mu$. The possible representations coming from the $U(1)$ 
part (the exact spectrum of $SU(4|2)$ representations in the U(1) theory) are 
given by
\begin{figure}[ht]
\vskip -0.3 cm
\centerline{\epsfysize=0.4truein \epsfbox{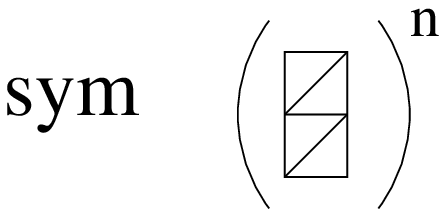}}
\vskip -0.3 cm
\label{symmn}
\end{figure} 
\newline
and this turns out to include exactly one of each allowed supertableau 
with an 
even number of boxes in each column. From now on, we 
will ignore this $U(1)$ part of the wavefunction and focus on the spectrum in 
the $SU(N)$ theory. 

\begin{figure}
\vskip -0.3 cm
\centerline{\epsfysize=6truein \epsfbox{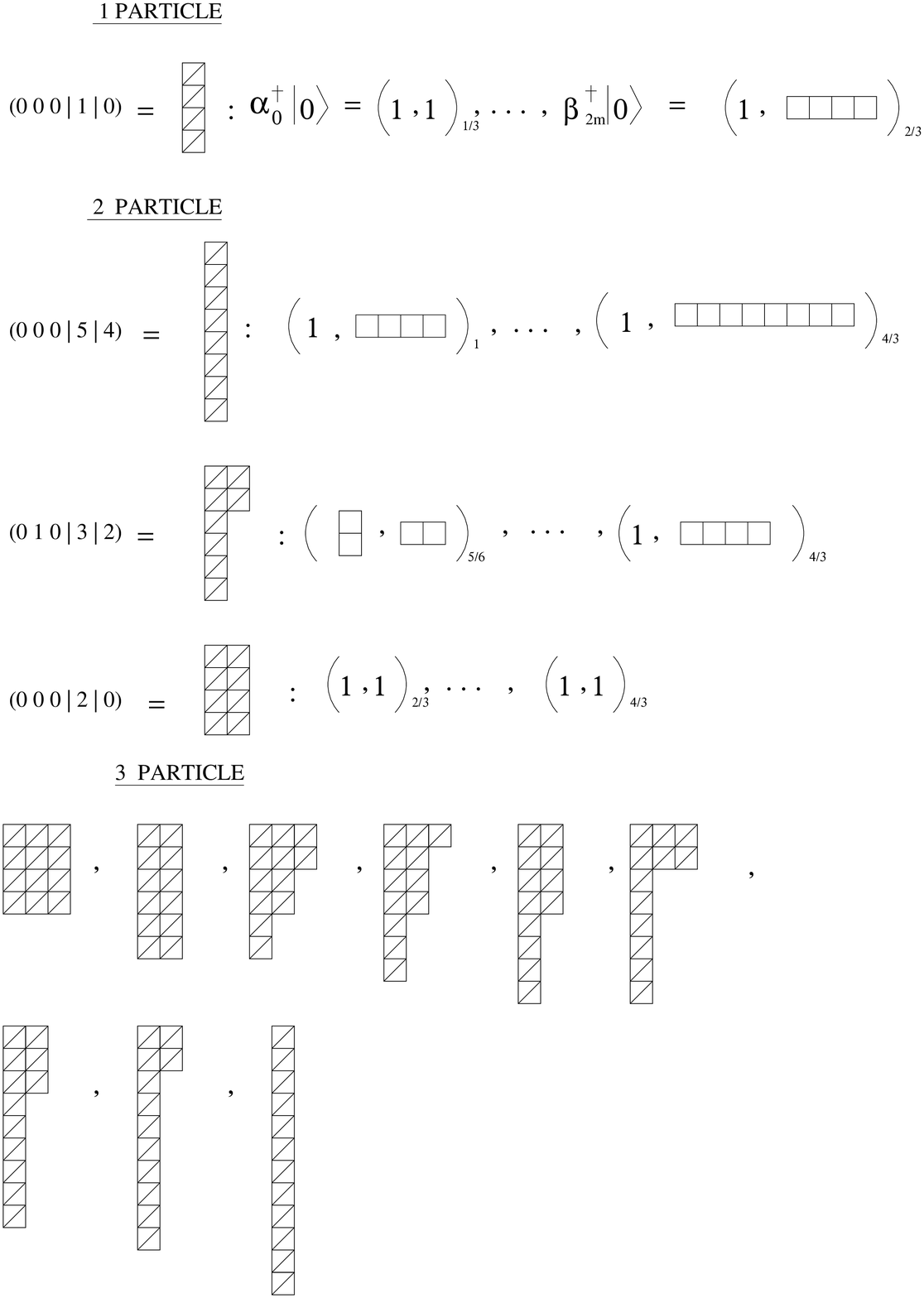}}
\vskip -0.3 cm
\caption{One, two and three particle multiplets of the $SU(2)$ theory
in the single-membrane vacuum.}
\label{XJ}
\end{figure}

For the simplest case of $SU(2)$, the representations are given by first factor 
in the tensor product of Figure \ref{nfour}. The supertableaux corresponding to 
the one, two, and three oscillator states are given in Figure \ref{XJ}. 

It is straightforward 
to determine the explicit oscillator expressions for these representations. For 
example, the highest energy states in the two-oscillator representations are the 
linear combinations of the states $\beta^\dagger_{2m} \beta^\dagger_{2\tilde{m}} 
| 0 \rangle$ with spin 4, 2, and 0, respectively. The other states in these 
representations may be determined explicitly by acting with the supercharge 
given in (\ref{QQ}).

For the general case of $SU(N)$, we have displayed all representations
corresponding to tableaux with ten or less boxes in Figure
\ref{8and10}, with labels denoting how they arise in the tensor
product of Figure \ref{nfour}.

\clearpage

\subsection{General vacua}

The analysis for general vacua is very similar to that of  the single membrane 
vacuum. A general vacuum corresponds to collections of $M_i$ coincident 
membranes at radii corresponding to momentum $N_i$. The oscillators for a 
general vacuum were described in section 5.3 of \cite{dsv} and are indicated in 
Table 2.

For each ordered pair $(N_i, N_j)$ of momenta (including the case $N_i = N_j$), 
we have a set of oscillators similar to those for the single membrane vacuum but 
which are $M_i \times M_j$ matrices and which have spins ranging from roughly 
$|N_i - N_j|/2$ to $(N_i + N_j)/2$. From the Table 2, is straightforward to 
show that these correspond to supertableaux shown in Figure \ref{Reduce}.

As explained in section 5.4 of \cite{dsv}, physical states satisfying the Gauss 
law constraint are obtained by acting with traces of arbitrary products of the 
matrix oscillators on the Fock-space vacuum. (Of course the product 
is 
required to give a square matrix in order to take a trace.) 

To work out the $SU(4|2)$ representations corresponding to these physical 
states, we may define super-oscillators
\[
(a^\dagger_{[I_1 \cdots I_{|N_i - N_j|+2}]})_{M_i \times M_j}, \dots , 
(a^\dagger_{[I_1 \cdots I_{(N_i + N_j)}]})_{M_i \times M_j}
\]
The set of representations may then be determined from the tensor product of 
tableaux corresponding to an allowed product of traces of these 
super-oscillators. One must be careful to keep only representations in 
the tensor 
product which survive any possible symmetrization, for example, the cyclic 
symmetry of the trace. Also, for finite $N$ one should keep in mind relations 
between traces of large numbers of matrices and products of lesser traces so as 
not to overcount representations.

\subsubsection*{Example: $X=0$ vacuum.}

\begin{figure}[ht]
\vskip -0.3 cm
\centerline{\epsfysize=0.5truein \epsfbox{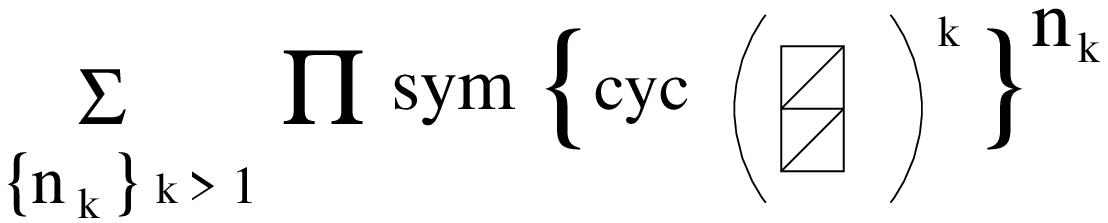}}
\caption{Spectrum of the $X=0$ vacuum for $N \to \infty$}
\label{symcyc}
\vskip -0.3 cm
\end{figure}

As an example, we consider the case of the $X=0$ vacuum, where $M_1 = N$  and 
$N_1 = 1$. Here, we have only a single matrix super-oscillator
\[
(a^\dagger_{[IJ]})_{N \times N}
\]
so the representations in the spectrum will be obtained by taking products of 
traces of powers of this oscillator. We will ignore the $U(1)$ part of the 
theory, so we assume that the oscillator $a^\dagger$ is traceless. In the large 
$N$ limit (where there are no trace relations), the complete spectrum of the 
$X=0$ vacuum is given by the tensor product of Figure \ref{symcyc} in
which $sym$ indicates a totally symmetrized tensor product $cyc$ indicates 
the cyclically symmetrized tensor product.
\newline

For finite $N$, one must take into account trace relations. As an example, for 
$SU(2)$ we may decompose traceless matrices as $A = A_i \sigma^i$. Then using
\[
\tr(AB) = 2 A_i B_i \qquad \qquad \tr(ABC) = 2i \epsilon^{ijk} A_i B_j C_k
\]
it is easy to see that the trace of any even number of matrices may be written 
in terms of products of traces of pairs of the matrices, while traces of any odd 
number of matrices may be written as products involving a single trace of three 
matrices and a number of traces of pairs of matrices. Thus, the spectrum for the 
$SU(2)$ case is obtained by restricting the sum in the tensor product
of Figure \ref{symcyc} to $n_3 =0,1$ and $n_i = 0$ for $i>3$.

The spectrum of representations involving three or less oscillators will be the 
same for any $N$, and this is depicted in Figure \ref{xzero}.

\section{Protected states in the matrix model}

Having specified the representations present in the spectrum at $\mu = \infty$ 
we would now like to see which states are protected from receiving energy 
corrections. 

\subsection{Exactly protected representations}

From the discussion in section 4, we concluded that all $SU(4|2)$ 
representations in Figure \ref{nosplit}, which include all doubly atypical 
representations, have no possibility of combining with other representations to 
form typical representations and therefore have energies that are completely 
protected from all perturbative and non-perturbative corrections. We
will now determine all representations in the matrix model spectrum of
the type in Figure \ref{nosplit}. 

We first note that all of the vacuum states, which lie in trivial 
representations of $SU(4|2)$ are doubly atypical and therefore are exact vacuum 
states quantum mechanically for all values of $N$ and $\mu > 0$.

\begin{figure}
\vskip -0.3 cm
\centerline{\epsfysize=0.5truein \epsfbox{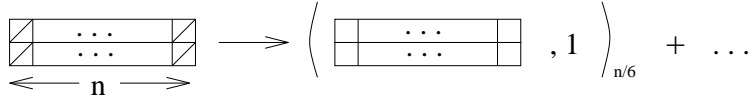}}
\caption{Doubly atypical representations in the matrix model spectrum}
\label{nby6}
\vskip 0.3 cm
\end{figure}

To see which other exactly protected representations exist in the matrix model 
spectrum, note that they may only arise from tensor products of oscillators in 
the representation
\begin{figure}[ht]
\vskip -0.3 cm
\centerline{\epsfysize=0.3truein \epsfbox{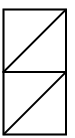}}
\vskip -0.3 cm
\end{figure} 
\newline
All other oscillators correspond to tableaux with heights of at 
least four 
boxes, and tensor products involving these can never result in a tableau with a 
height of two or three boxes. The only representations in Figure \ref{nosplit}
that can arise from tensor products of this two-box representation are
the singly atypical representations
\begin{figure}[ht]
\vskip -0.3 cm
\centerline{\epsfysize=0.4truein \epsfbox{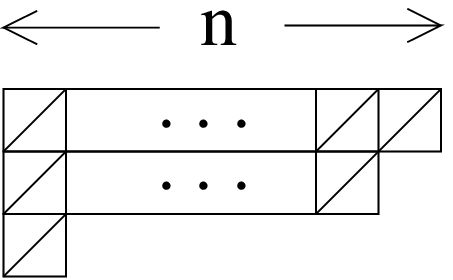}}
\vskip -0.3 cm
\label{single}
\end{figure} 
\newline
which we will discuss last, and the doubly atypical representations
shown in Figure \ref{nby6} arising from products of $n$
super-oscillators $a_{IJ}$ which are completely 
symmetrized. Note that the lowest 
energy $SU(4) \times SU(2)$ representation in the decomposition of
these doubly atypical
representations has trivial $SU(2)$ part and an $SU(4)$ 
representation which is equivalent to the completely symmetric,
traceless 
$n$-index tensor representation of $SO(6)$. Recalling that the lowest-energy 
component of the two-box representation is always an $SO(6)$ vector oscillator  
with energy $1/6$, we see that the doubly atypical representations are precisely 
those built upon states with $n$ $SU(2)$-singlet $SO(6)$ oscillators whose 
vector indices are contracted with a completely symmetric traceless tensor 
(with 
arbitrary $SU(N)$ trace structure). 

For example, about the $X=0$ vacuum, we have a single $N \times N$ matrix 
$SO(6)$ vector oscillator $X^a$ with spin 0, so the doubly atypical 
representations are those built upon the states\footnote{These have a
structure identical to the chiral primary operators of $N=4$ SYM
theory. This is probably since $SU(4|2)$ is a subgroup of the superconformal
algebra $SU(2,2|4)$ governing that theory, so we expect that the chiral
primary operators of AdS/CFT lie in doubly atypical representations of
this subgroup.} 
\beas
&&S_{ab} \tr(A_a^\dagger A_b^\dagger) |0 \rangle , \qquad S_{abc} 
\tr(A_a^\dagger 
A_b^\dagger A_c^\dagger) |0 \rangle, \qquad  S_{abcd} \tr(A_a^\dagger 
A_b^\dagger A_c^\dagger A_d^\dagger) |0 \rangle, \cr
&& \qquad \qquad S_{abcd} \tr(A_a^\dagger A_b^\dagger) \tr(A_c^\dagger
A_d^\dagger) |0 \rangle, 
\dots
\eeas
where $A_a^\dagger$ is the matrix creation operator associated with the 
oscillator $X^a$ and the tensors $S$ are completely symmetric and traceless.
Note that some of the states in this series will not be independent for finite 
values of $N$, as discussed above. As an explicit check, we have computed 
the 
leading perturbative correction (at second order in perturbation theory) for the 
first state on this list in appendix B and found that it indeed vanishes.

For the general vacuum with $M_i$ spheres of radius $N_i$, we have one $M_i 
\times M_i$ matrix spinless SO(6) vector oscillator $(A_a^i)^\dagger$ for each 
$N_i$, so the doubly atypical representations about a general vacuum will be 
those built upon states such as
\[
S_{abcdef} \tr((A^1_a)^\dagger) \tr((A^1_b)^\dagger (A^1_c)^\dagger 
(A^1_d)^\dagger ) \tr((A^2_e)^\dagger (A^2_f)^\dagger) |0 \rangle
\]
Note that we may have traces of single oscillators for a general vacuum 
since only the combination $\sum_i \tr(A^i_a)$ corresponds to the $U(1)$ 
part.

Finally, we turn to the singly atypical representations with three 
rows, depicted in the 
previous page. 
The lowest energy states in these representations, again with energy $h
=n/6$, are in a
trivial representation of $SU(2)$ and an $SU(4)$ representation
$(1,n-2,1)$ corresponding to the $n$-index tensor of $SO(6)$ with all but one 
of 
the
indices symmetrized. These
must be states formed from $n$ spinless $SO(6)$ creation operators with
$n-1$ of the indices totally symmetrized and a single pair of
antisymmetrized indices. Because of the antisymmetric pair, this will
be non-vanishing only if at least two different types of spinless
$SO(6)$ oscillators are present (i.e. the state must involve center of mass
motions of two collections of membranes at different radii). For
example, the lowest energy states in the representation for $n=2$ take the form
\[
\tr((A^1_{[a})^\dagger) \tr((A^2_{b]})^\dagger |0 \rangle 
\]
where $A^1$ and $A^2$ must be distinct square matrix oscillators of
different size.

Thus we find that in addition to the vacuum states, 
the matrix model spectrum for $\mu = \infty$ contains
infinite towers of  representations of the type shown in Figure 
\ref{nosplit}. 
Based on the representation theory of $SU(4|2)$, we may conclude that all of 
these representations are preserved for any value of $\mu$ and that their 
energies receive no perturbative or non-perturbative corrections. For any given 
vacuum, the spectrum of these representations has a well defined large 
$N$ limit (obtained by ignoring any trace relations) so we may conclude that 
these states are present as quantum states in the exact spectrum of
$M$-theory on the pp-wave background.

\subsection{Singly atypical representations}

In addition to doubly atypical representations, the spectrum of the
matrix model contains infinite towers of singly atypical
representations, for example all the single particle multiplets about the
single membrane vacuum. As we
discussed in section 4, these are protected
from receiving energy shifts except in cases where they combine with
another atypical representation to form a typical representation. 

A given atypical representation may combine with at most two other atypical
representations, obtained either by adding a box to each of the first three
rows of the supertableau and removing a box from the first column, or
removing a box from each of the first three rows and adding a box to
the first column, as shown in Figure \ref{combine}. In cases where neither 
of
these atypical representations appear in the spectrum at $\mu =
\infty$, we may conclude that the energy of the original atypical
representation does not change as we vary $\mu$ away from
$\infty$.\footnote{If one of these representations does appear, then
the energy may or may not shift.}

A weaker assertion may be made when the neither of these possible
atypical representations appear as states about the same vacuum as the
original atypical representation but possibly appear as states about
another vacuum. In this case, we expect that the original atypical
multiplet is protected from receiving any perturbative corrections, but
non-perturbatively there may be an energy shift since the two
multiplets about the different vacua could possibly mix and combine as
$\mu$ is decreased away from $\infty$.

To start with, we will see which states are protected perturbatively
for the single-membrane and $X=0$ vacua and then turn to the stronger
condition of non-perturbative protection.

\begin{figure}
\vskip -0.3 cm
\centerline{\epsfysize=3.5truein \epsfbox{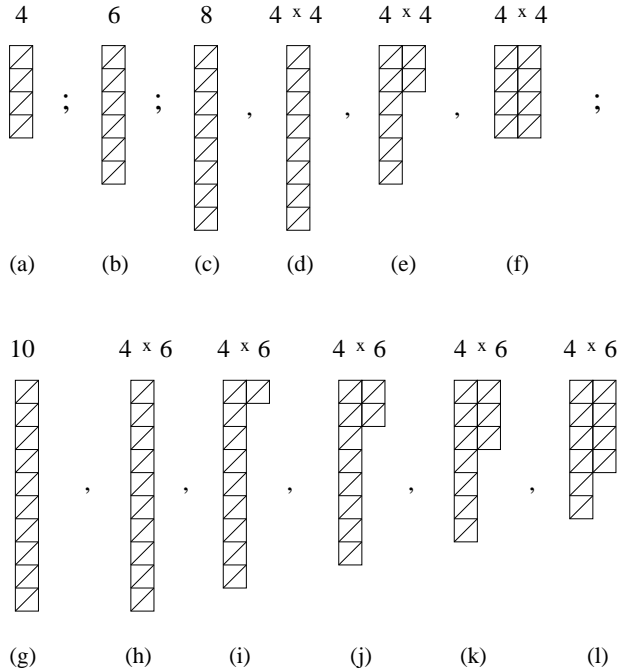}}
\vskip -0.3 cm
\caption{ $SU(4|2)$ multiplets with ten or less boxes for the $SU(N)$
single membrane vacuum.}
\label{8and10}
\end{figure}

\subsubsection*{Perturbative energy shifts: single-membrane vacuum}
 
We begin by considering the set of atypical multiplets about the single
membrane vacuum, starting with the simplest case of $SU(2)$ (i.e. $N=2$). 

In this case, it is simple to argue that all atypical multiplets, such
as the seven multiplets in Figure \ref{XJ} with less than four boxes in
the second column, are
protected from receiving any perturbative energy shifts. To see this,
note that the spectrum of representations about the single-membrane vacuum for
$SU(2)$ is generated by tensor products of a supertableau with four
boxes. Thus, the supertableaux corresponding to all representations in
the spectrum will have multiples of 4 boxes. But from Figure \ref{combine}, 
we see that the number of boxes for pairs of atypical representations
that can combine always differ by 2.\footnote{It is important to note
that while the equivalence between supertableaux given in Eq.(\ref{equ}) 
equates
tableaux whose number of boxes can differ by 2, only typical
representations may have two inequivalent tableaux. } Hence, none of the 
atypical
multiplets in the spectrum can combine and therefore all have energies
which are protected from any perturbative corrections.

We now turn to the general case of $SU(N)$. The multiplets with
tableaux containing up to 10 boxes are displayed in Figure \ref{8and10}. It is
easy to check that the atypical tableaux labelled by {\bf a}, {\bf b}, and {\bf 
e}
cannot receive perturbative energy corrections since the other
representations with which they could combine do not appear in the
perturbative spectrum. As a check, we note that the lowest energy
state in the multiplet {\bf a} is 
\[
a^\dagger_{00} |0 \rangle
\]
whose energy shift was computed to second order in perturbation
theory in \cite{dsv} and found to be zero. 

For the equivalent multiplets {\bf c} and {\bf d} , there is a
possibility of combining with the multiplet {\bf k} since these
multiplets are related as in Figure \ref{combine}. Thus, we may only conclude
that one linear combination of the multiplets  {\bf c} and {\bf d} 
is protected from receiving an energy shift. On the other hand, we
have found that the leading perturbative energy shifts for all three
of these multiplets
is zero.\footnote{This may be shown without a full calculation. First,
the multiplet {\bf d} is protected for $SU(2)$; and
for $SU(N)$, one finds that all diagrams contributing to the
leading-order shift are proportional to those for $SU(2)$ and
therefore cancel. The multiplet {\bf k} is therefore protected to
leading order for $SU(3)$ (the lowest $N$ for which it appears) and we
find that all contributions to the leading-order energy shift of {\bf k} for 
SU(N)
are proportional to those for $SU(3)$ and therefore cancel. The
multiplet {\bf c} is therefore also protected at leading order since
it would have to have the same shift as {\bf k}.}
Thus, despite the possibility for an energy shift, we do not find one,
at least to leading order. 

By considering tableaux with 12 boxes one may show that the multiplets
{\bf i} and {\bf j} are protected, while the multiplets {\bf g} and
{\bf h} have the possibility of combining with other multiplets and
shifting. Again, explicit calculation indicates that these multiplets
do not shift to leading order.

To conclude our discussion of perturbatively protected states for the
single membrane vacuum, we note that the following infinite tower of
multiplets are protected from receiving perturbative energy corrections: 
\begin{figure}[ht]
\vskip -0.3 cm
\centerline{\epsfysize=0.8truein \epsfbox{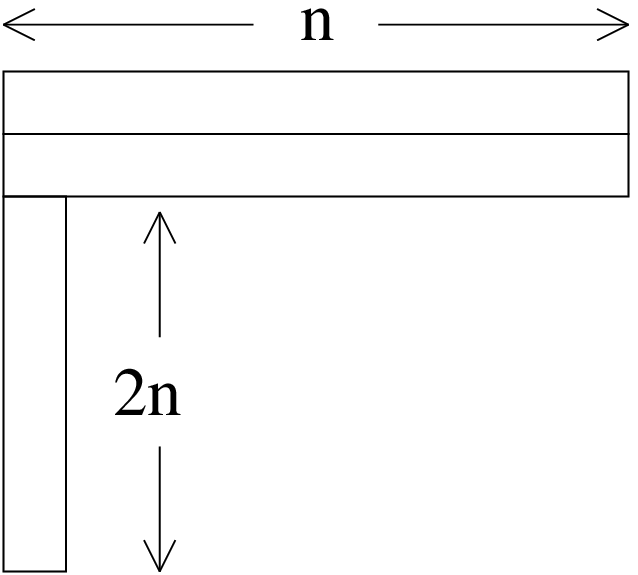}}
\vskip -0.3 cm
\end{figure} 
\newline
To see that these are protected, note that for a given $n$, this
multiplet could only combine with the multiplet corresponding to the tableau
\begin{figure}[ht]
\vskip -0.3 cm
\centerline{\epsfysize=0.8truein \epsfbox{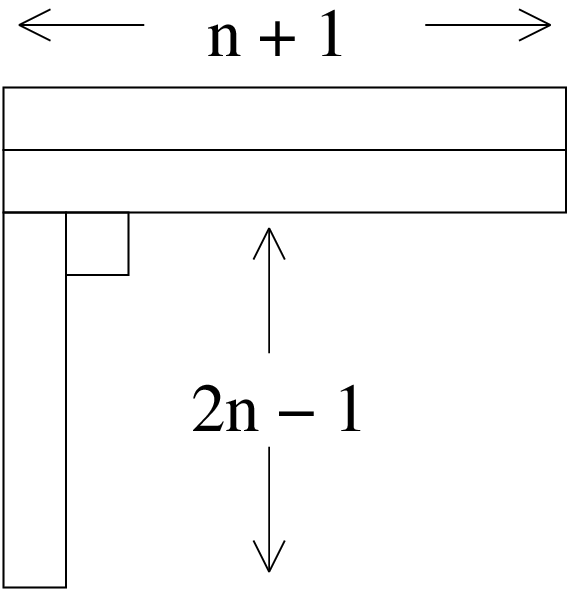}}
\vskip -0.3 cm
\end{figure}
\newline 
which has $4n+2$ boxes and $n+1$ columns. However, the maximum number
of columns in a tableau with $4n+2$ boxes (for the single-membrane
vacuum) is $n$, since such a tableau can arise from the tensor
product of at most $n$ oscillator representations ($n-1$ 4-box
representations and a single 6-box representation). 

To see explicitly which states these protected multiplets correspond
to, note that one $SU(4) \times SU(2)$ representation in the
decomposition (for which we associate all boxes in the first two rows
with $SU(4)$) is the $n$-index traceless symmetric tensor of $SO(6)$
with spin $n$ and energy $h = n/2$. This is the multiplet
corresponding to a set of $n$ spin 1 $SO(6)$ oscillators with spins
aligned and $SO(6)$ indices completely symmetrized, for example
\[
S_{a_1 \cdots a_n} (a^{a_1}_{11})^\dagger \cdots (a^{a_1}_{11})^\dagger |0 
\rangle
\]
These have a structure which is very similar to the lowest energy
states of the doubly atypical representations discussed above, except
that the oscillators now carry spin. It would be interesting to
determine if similar states formed out of higher spin $SO(6)$
oscillators are also protected.

\subsubsection*{Perturbative energy shifts: X=0 vacuum}

\begin{figure}
\vskip -0.3 cm
\centerline{\epsfysize=2.5truein \epsfbox{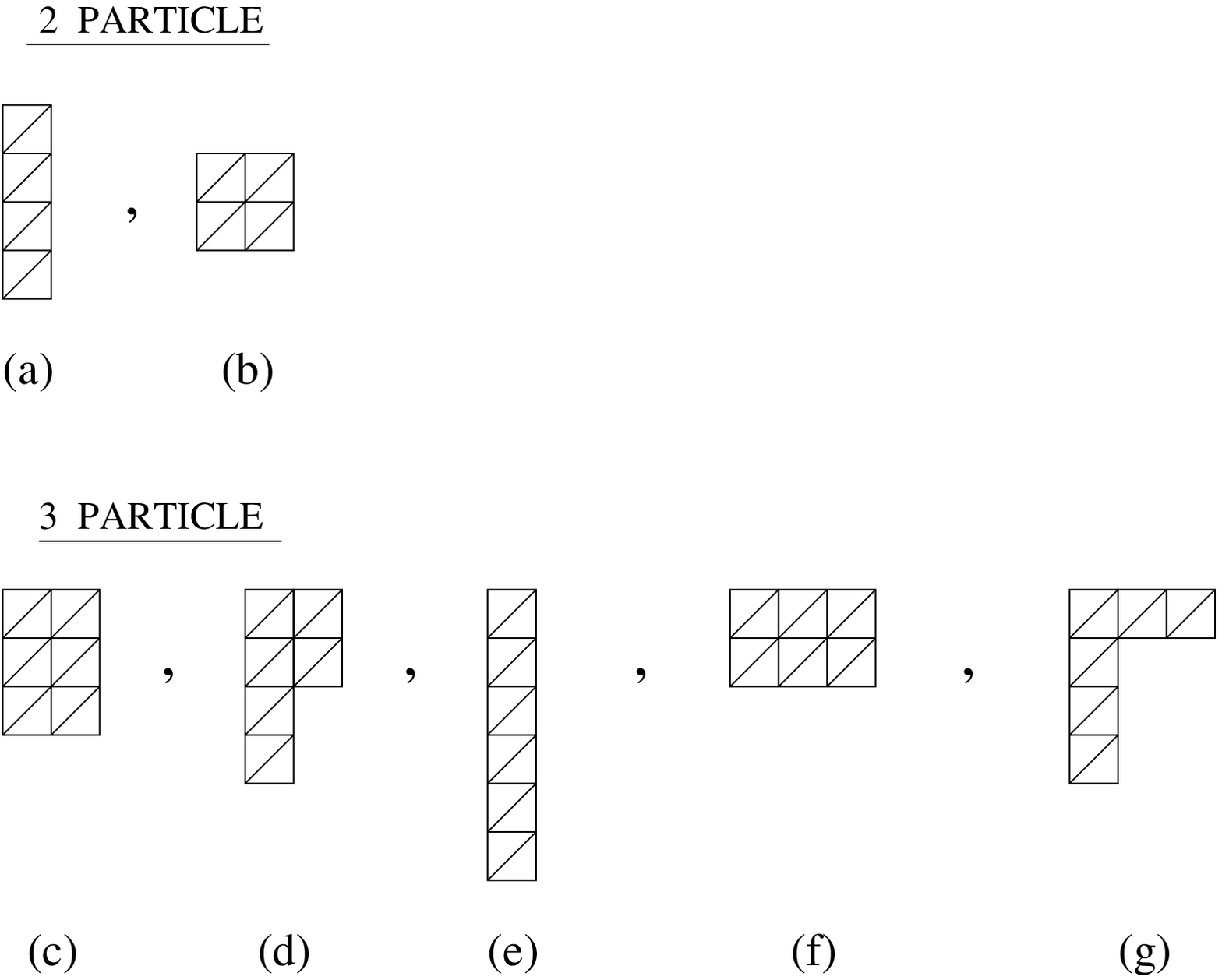}}
\vskip -0.3 cm
\caption{Two and three oscillator multiplets for the $X=0$ vacuum}
\label{xzero}
\end{figure}

We now consider the low-energy multiplets for the $X=0$ vacuum, depicted in 
Figure \ref{xzero}. Among the two-oscillator states, the multiplet {\bf b} is 
the 
protected doubly atypical multiplet that we have already considered (whose 
lowest 
energy state is $S_{ab} \tr(A_a^\dagger A_b^\dagger)|0 \rangle$). In addition, 
we have a singly atypical multiplet {\bf a} with lowest energy state 
\be
\label{state}
\tr(A_a^\dagger A_a^\dagger)|0 \rangle \; .
\ee
From Figure \ref{combine}, we note that this multiplet has the possibility of 
combining 
with the atypical multiplet {\bf c} in the three-oscillator sector whose lowest 
energy states are given by 
\[
A_{abc} \tr(A_a^\dagger A_b^\dagger A_c^\dagger)|0 \rangle \; 
\]
where $A_{abc}$ is a completely antisymmetric anti-self-dual tensor of $SO(6)$ 
(in the (0,0,2) representation of $SU(4)$). To see whether this occurs, we may 
look for an energy correction to the state (\ref{state}) at leading order in 
perturbation theory. In fact, we have done this computation previously in 
\cite{dsv} and again in appendix B and found that there is in fact a positive 
shift in the energy. Thus, we may conclude that the representations {\bf a} and 
{\bf c} do combine to form a typical multiplet as $\mu$ is decreased from 
$\infty$.

By considering the multiplets containing 8 boxes, we find that the singly 
atypical representations {\bf d} and {\bf e} are perturbatively protected since 
the representations that they could combine with are not in the spectrum. The 
8-box atypical representation that can combine with the representation {\bf g} is 
in the spectrum, so it is possible that {\bf g} receives an energy shift. 
Finally, the representation {\bf f} is one of the protected doubly
atypical multiplets considered above. 

\subsubsection*{Non-perturbatively protected states}

In the previous subsections, we have shown that certain singly atypical 
representations are protected from receiving any perturbative energy corrections 
due to the absence of representations above the same vacuum with which they can 
combine. However, it is possible that pairs of atypical representations above 
different vacua could combine, leading to a non-perturbative shift in the energy 
even for states that are perturbatively protected to all orders.

By applying our group theory reasoning to the total spectrum including states 
above all vacua we would now like to see when this non-perturbative mixing 
might 
occur. In the process, we will also be able to strengthen our assertions for 
the protection of certain states from perturbative to non-perturbative 
statements.

For simplicity, we will focus on $SU(2)$ whose only vacua are the $X=0$ vacuum 
and 
the single-membrane vacuum we have already considered. The low energy multiplets 
for these two vacua are depicted in Figures \ref{XJ} and \ref{xzero}. 

Among the four-box representations, the doubly atypical
representation {\bf b} above the $X=0$ vacuum is certainly protected 
non-perturbatively, as discussed above. There are two 
other four-box representations, the multiplet {\bf a} in the $X=0$ 
vacuum and the identical multiplet in Figure \ref{XJ} for the single-membrane
vacuum. Among the six-box multiplets, 
there is a single copy of the representation with which these can combine, so we 
may conclude that one combination of these representations must be protected 
non-perturbatively (we have already seen that one combination gets a shift). 

Among the remaining six-box representations about $X=0$, {\bf f} is doubly 
atypical and hence protected non-perturbatively. As for  the perturbatively 
protected multiplets {\bf d} and {\bf e}, we may now conclude 
that they are also protected 
non-perturbatively, since there are no single-membrane multiplets with which 
these could combine. 

As a final example, we note that the perturbatively protected representation 
indicated by the second supertableau of Figure \ref{XJ} has the
possibility of a non-perturbative 
energy shift, since the $X=0$ spectrum contains two copies of the 10-box 
representation with which this can combine.\footnote{These are five oscillator 
states and therefore don't appear in the Figure \ref{xzero}, but may be 
obtained from the general formula in Figure \ref{symcyc}.}

Before concluding this section, we emphasise that even for states which are
protected from receiving an energy shift (perturbatively or
non-perturbatively) as $\mu$ is varied from $\mu=\infty$, there is a 
possibility 
that there could be a shift beginning at some finite value of $\mu$, as 
explained in section 4.2. In certain cases, however, it is posible to argue 
using the supersymmetric indices defined in Eq.(\ref{index}), that a 
multiplet (other than those of Figure \ref{nosplit}) must be present in the 
spectrum for all values of $\mu$. 

As an example, we continue to focus on the $SU(2)$ theory and consider the 
simplest 
non-trivial chain of the type depicted in Figure \ref{chains}, namely the chain 
with two elements corresponding (from left to right) to tableaux {\bf c} and 
{\bf a} of Figure \ref{xzero}.  It is a trivial matter to compute the 
corresponding 
index from 
our knowlege of the spectrum at $\mu = \infty$, 
\[
{\cal I}(0,0,2) = {\cal N}(0,0,2|0|0) - {\cal N}(0,0,0|1|0) = 1 - 2 = -1 
\]
Since ${\cal N}(0,0,2|0|0)$ must be non-negative for any value of $\mu$ we may 
conclude that at least one copy of the representation with tableau as in Figure 
\ref{xzero}a must be in the spectrum for all values of $\mu$.

Thus, in various cases we have demonstrated arguments that certain multiplets 
are protected perturbatively to leading order, perturbatively to all orders, 
non-perturbatively near $\mu = \infty$, and/or non-perturbatively for all $\mu 
> 
0$.
We next consider a class of non-protected states and show that even these 
benefit from certain supersymmetric cancellations.

\subsection{Cancellations for typical states}

In \cite{dsv}, we noted that the leading order vanishing energy shift for the 
state
$a^\dagger_{00} |0 \rangle$ (which we have seen is protected) leads to
cancellations in the leading order energy shifts for the non-protected states
$(a^\dagger_{00})^n |0 \rangle$ such that the non-vanishing
contribution to the energy shift gives a finite result in the large
$N$ limit. In this section, we will extend this result by showing that
the leading perturbative corrections
(which come at second order in perturbation theory) to the 
energies of all states about the single-membrane vacuum have finite large 
$N$ limits. 

\begin{figure}
\vskip -0.3 cm
\centerline{\epsfysize=0.8truein \epsfbox{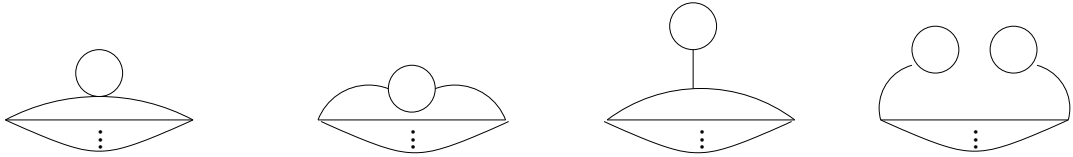}}
\vskip -0.3 cm
\caption{Loop diagrams contributing to the leading energy shifts}
\label{Feynman}
\end{figure}

We first recall from section 6.1 of \cite{dsv} that when the 
interaction Hamiltonian is expanded about the single-membrane vacuum in terms 
of 
canonically normalized oscillators, the coupling that appears is 
\[
g = \left({R \over \mu N}\right)^{3 \over 2}
\]
which is fixed in the large $N$ limit defining M-theory. Thus, the only 
possible
divergent $N$ dependence in the leading perturbative energy shifts is
from the sum over intermediate states in loops. As shown in Figure 
\ref{Feynman}, the only diagrams which contain loops are those for which 
all 
but one of the creation operators from the initial state are contracted with
annihilation operators from the final state.\footnote{Diagrams for
which there is a single loop connected only to the states on the left
or on the right cannot appear since in this case, the left and right
states would not have the same tree-level energy.} Up to combinatorial
factors, these diagrams therefore reduce to the diagrams
contributing to the energy shifts of single oscillator
states.
It follows that all states will have finite leading energy shifts in
the large $N$ limit as long as all of the single particle states do. 
\newline
The rest of argument proceeds by induction. We have already seen that all
single-particle states in the four-box representation of $SU(4|2)$ are
protected and therefore have zero energy shifts in perturbation
theory. Now, suppose we have proven that all single particle states up
to and including those in the $2n$-box representation of $SU(4|2)$
have finite energy shifts at leading order. To show that the single
particle states
in the $(2n+2)$-box representation must have a leading energy shift
which is finite in the large $N$ limit, note that these states are in an
atypical multiplet which can receive an energy shift only if it
combines with another atypical multiplet and becomes typical. From
Figure \ref{combine}, the only atypical multiplet with which the single-column
$(2n+2)$-box multiplet can combine is one with $2n+1$ boxes in the first
column and three boxes in the second column. However, any individual
oscillators contributing to the states in this multiplet must be from
$SU(4|2)$ representations with at most $2n$ boxes, so the leading 
perturbative energy shift for such a representation must be finite in
the large $N$ limit. It the two atypical multiplets do combine to
form a typical multiplet and thus receive an energy shift, the shift
for the states in the $(2n+2)$-box multiple must be identical to the
shift for the states in the $(2n+1,3)$-box multiplet, and therefore
must also be finite in the large $N$ limit. This completes the inductive
proof.

Hence, at least to the leading non-trivial order, perturbative
corrections to the energies of states about the single-membrane vacuum
are completely well-behaved in the large $N$ limit. 
We will discuss this result further in section 8.

\section{Relation between atypical representations \\ and BPS states}

In this paper, we have seen that the spectrum of the pp-wave matrix model must 
fall into representations of the superalgebra $SU(4|2)$. This superalgbra has 
special multiplets which are called atypical with the property that there are no 
nearby multiplets with the same $SU(4) \times SU(2)$ state content but different 
energy. Atypical multiplets have fewer states than the typical multiplets with 
nearby highest weights. Further, the energies of states in atypical multiplets 
are protected unless the atypical multiplet combines with another atypical 
multiplet to form a typical multiplet.

It is clear that many of the properties of atypical multiplets are the same as 
properties usually associated with BPS multiplets in more familiar
supersymmetry algebras. In the 
usual case, a BPS state is defined to have the property that it is annihilated 
by one or more of the hermitian supersymmetry generators, and all states in a 
BPS multiplet share this property. In \cite{dsv}, we showed that the matrix 
model contains infinite towers of states which are BPS in this usual sense. 
It is natural to guess that these BPS states will be associated with atypical 
multiplets, however, the precise connection between BPS states and atypical 
multiplets certainly requires clarification, which we now provide.

We will show that all atypical multiplets contain BPS states and all BPS states 
belong to atypical multiplets. However, not all states of an atypical multiplet 
are BPS. Indeed, there are atypical states carrying no charges at all that still 
have protected energies!

\begin{figure}
\vskip -0.3 cm
\centerline{\epsfysize=0.9truein \epsfbox{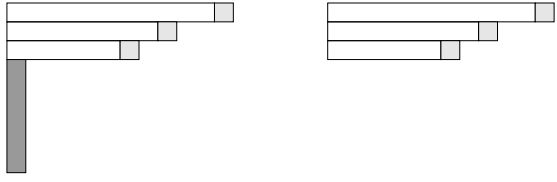}}
\vskip -0.3 cm
\caption{BPS states from atypical supermultiplets}
\label{grey}
\end{figure} 

We begin by recalling the discussion of BPS states in \cite{dsv}. Starting from 
the relation
\be
\label{Q,Q}
\{ Q_\alpha, Q_\beta \} = 2 \delta_{\alpha \beta} H - { \mu \over 3} 
(\gamma^{123} \gamma^{ij})_{\alpha \beta} M^{ij} + { \mu \over 6} (\gamma^{123} 
\gamma^{ab})_{\alpha \beta} M^{ab}
\ee
and choosing a set of Cartan generators $M^{12}$, $M^{45}$, $M^{67}$, and 
$M^{89}$ for $SO(6) \times SO(3)$, we showed that the eigenvalues of 
$\langle \psi | \{ Q_\alpha , Q_\beta \} | \psi \rangle $ for states
in any $SO(6) \times SO(3)$ multiplet are given by two copies of the set 
\[
\Delta = H + \epsilon_1 {\mu \over 3} M^{12} + \epsilon_2 {\mu \over 6} M^{45} 
+ \epsilon_3 {\mu \over 6} M^{67} - \epsilon_2 \epsilon_3 {\mu \over 6} M^{89}
\]
where $\epsilon_i = \pm 1$ are chosen independently and $M$'s are
eigenvalues of the Cartan generators for the heighest weight state in the 
multiplet.
Thus, BPS $SO(6) \times SO(3)$ multiplets are those for which
\be
\label{BPScond}
\lambda \cdot (1, \epsilon_1, \epsilon_2, \epsilon_3, -\epsilon_2 \epsilon_3) = 
0
\ee
for one or more choices of $\epsilon_i$, where
\[
\lambda \equiv (3h, M^{12}, {1 \over 2} M^{45}, {1 \over 2} M^{67}, {1 \over 2} 
M^{89}) \; .
\] 

To relate this to our discussion in the rest of the paper, we should first 
determine the relation between this choice of Cartan generators and the Dynkin 
basis used in the rest of the paper. By explicitly writing the rotation 
generators in $SU(4) \times SU(2)$ notation, we find 
\bea
M^{12} &=& diag(0,0,0,0,{1 \over 2}, -{1 \over 2}) = {1 \over 2} H_5\cr
M^{45} &=& diag(1,1,-1,-1,0,0) = {1 \over 2} (H_1 + 2 H_2 + H_3)\cr
M^{67} &=& diag(1,-1,1,-1,0,0) = {1 \over 2} (H_1 + H_3)\cr
M^{89} &=& diag(-1,1,1,-1,0,0) = {1 \over 2} (H_3 - H_1)\cr
3h &=& diag( {1 \over 4}, {1 \over 4},{1 \over 4},{1 \over 4},{1 \over 2}, {1 
\over 2}) = {1 \over 4} H_1 + {1 \over 2} H_2 + {3 \over 4} H_3 + H_4 - {1 \over 
2} H_5
\label{rel2}
\eea
Now, starting with any $SU(4|2)$ representation, we can consider any $SU(4) 
\times SU(2)$ representation in the decomposition, determine its
highest weight in the Dynkin basis, calculate $\lambda$ using
Eq.(\ref{rel2}), and check whether the condition 
(\ref{BPScond}) is satisfied.

We first note that no typical representation can contain a BPS state, since the 
nearby typical representation with smaller $a_4$ would have a state with 
identical charges but smaller energy and therefore violate the BPS bound.

We next consider general atypical representations as depicted in Figure 
\ref{grey}. It is straightforward to verify that any $SU(4) \times SU(2)$ 
representation  
obtained by assigning all of the dark gray boxes (for supertableau with four or 
more rows) and either 0, 1, 2, or 3 of 
the light gray boxes to a fully symmetric $SU(2)$ representation satisfy the 
condition (\ref{BPScond}) for at least one choice of $\epsilon_i$. All 
other representations in the decomposition do not. For $SU(2)$ non-singlet 
multiplets obtained in this way, the number of supercharges preserved 
(twice the 
number of choices of $\epsilon_i$ for which (\ref{BPScond}) is satisfied) 
is 8 minus
 twice the number of rows in the $SU(4)$ tableau. For BPS $SU(2)$
singlet multiplets, which arise only as the lowest energy multiplets
for supertableaux with three or less rows (shown on the right in
Figure \ref{grey}), the number of preserved supercharges is equal to 16
\begin{figure}
\centerline{\epsfysize=1.8truein \epsfbox{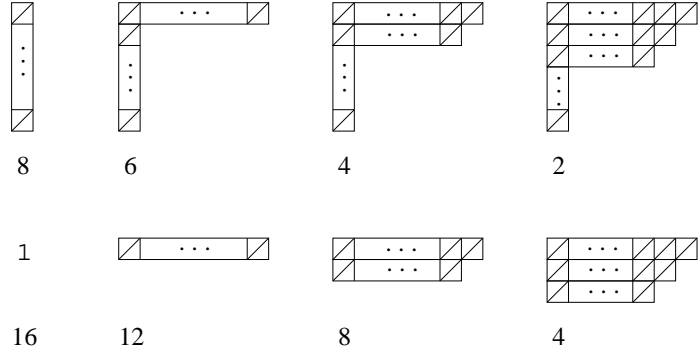}}
\caption{Maximum number of
supersymmetries preserved among
states of the atypical supermultiplets shown.}
\label{8642}
\end{figure} 
minus four times the number of rows in the $SU(4)$
tableau.\footnote{The increased number of preserved supercharges for the 
special
case of tableaux with less than four rows was not noted in the
original version of this paper. We realized that this case must be
treated separately upon reading the work of Kim and Park \cite{KP}.}
The fraction of 32 supersymetries preserved for the various BPS $SU(4)
\times SU(2)$ multiplets may be summarized as:
\begin{figure}[ht]
\centerline{\epsfysize=1.35truein \epsfbox{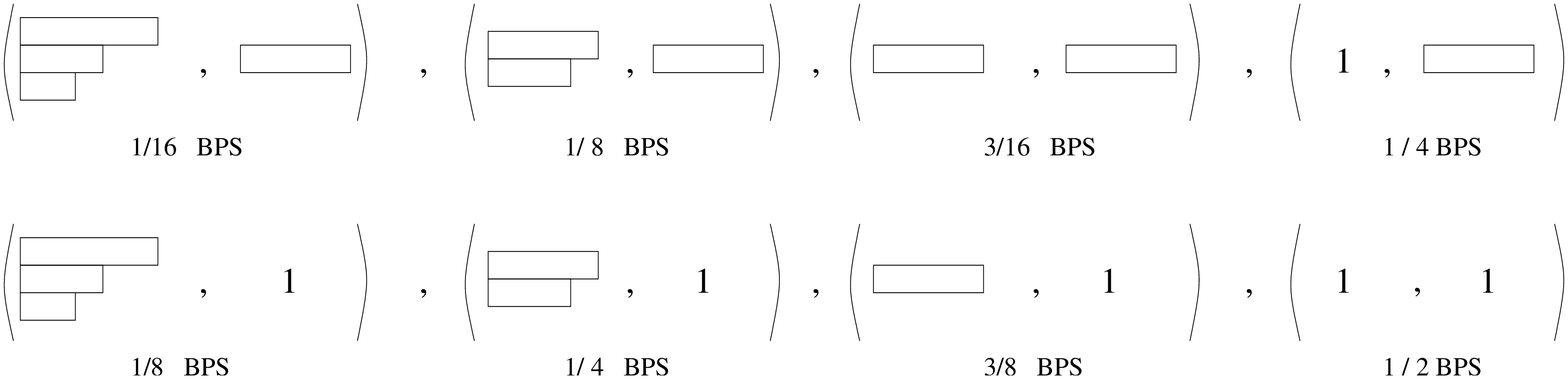}}
\label{bpsstates}
\end{figure} 
\newline
Note that the representations shown will only be BPS if they have the appropriate 
energy, which is true only if they arise from
an atypical multiplet in the manner depicted in Figure \ref{grey}. 

For a given atypical $SU(4|2)$ representation, we may then determine the 
$SU(4) \times SU(2)$ multiplet in its decomposition for which the
largest number of supercharges is preserved. This maximum number is indicated
in Figure \ref{8642} for all possible atypical mutiplets. In cases
where more than one of the tableaux apply (e.g. for the supertableaux
with three vertical boxes which could be described both by the first and the last
tableaux in Figure \ref{8642}) the maximum number of supercharges preserved is
given by the largest number among the applicable tableaux. It is interesting to 
note that the tableaux in the lower row with ``doubled'' supersymmetry are those 
corresponding to the isolated point in Figure \ref{reps} with $a_4=a_5=0$. 
Finally, we note that the physical spectrum of the matrix model contains examples
of all of these types of representations except for those with 12 supercharges
preserved. 

We emphasize that even atypical multiplets containing states preserving 8 
supercharges may also contain non-BPS states. A simple example is provided by 
the single oscillator states about the single-membrane vacuum of the $SU(2)$ 
theory. These form a single atypical multiplet which includes states preserving 
eight ($\beta_2$), six ($\eta_{3 \over 2}$), four ($x_1$), two ($\chi_{1 
\over 2}$), and 0 ($\alpha_0$) supercharges. Thus, while $\alpha_0$ is non-BPS 
and carries no charges whatsoever, it is part of an atypical multiplet and (as we 
have seen) its energy is protected.

\section{Discussion}

In this paper, we continued our analysis of the matrix model for M-theory on the 
maximally supersymmetric pp-wave background by exploring the physical 
consequences of the symmetry algebra. From its bosonic subalgebra $SU(4)\times 
SU(2)\times U(1)_H$, we identified the symmetry algebra of the interacting 
$SU(N)$ part of the theory to be the basic classical Lie superalgebra 
$SU(4|2)$. Using the work of Kac \cite{kac}, Bars et.al. \cite{bars} and 
Jakobsen \cite{jakobsen}, we described the complete set of physically allowable 
representations of $SU(4|2)$. Among these are the typical multiplets, which lie 
on one-parameter families of representations differing only by their energy, and 
the atypical multiplets, for which there are no nearby multiplets with the same 
$SU(4) \times SU(2)$ state content but different energy. We argued that states 
in atypical multiplets can only receive energy corrections if two atypical 
multiplets combine into a typical multiplet, and found the complete set of 
multiplet pairs for which this is possible. Certain multiplets, including those 
known as doubly atypical do not appear in any of the pairs, and therefore 
are completely protected. Additional exactly protected quantities are provided 
by the supersymmetric indices given in  Eq.(\ref{index}).

Equipped with this knowledge of the $SU(4|2)$ representation theory, we turned 
to 
the actual spectrum of the matrix model for $\mu = \infty$, identified the 
complete set of $SU(4|2)$ multiplets present, and used the representation theory 
reasoning to deduce which multiplets are protected as $\mu$ is made finite. We 
explicitly identified all doubly atypical (and therefore exactly protected) 
multiplets in the spectrum, and found that these include the vacuum states as 
well as infinite towers of excited states above the various vacua. Among the 
remaining atypical multiplets, we found some which are protected (either 
perturbatively or non-perturbatively) due to the absence of the complementary 
representations with which they could combine (either above the same vacuum or 
in the entire spectrum), and some which do combine and receive energy shifts (as 
verified by explicit perturbative calculation). Finally, we showed that the 
representation theory implies cancellations in the leading perturbative energy 
shifts for all states (typical and atypical) above the single-membrane vacuum, 
leaving a result that is finite in the large $N$ limit.

By identifying protected multiplets in the matrix model for $\mu = \infty$ we 
have provided non-trivial information about the spectrum of the matrix model for 
all values of $\mu$ including the regime where perturbation theory is 
inapplicable. In particular, since the exactly protected doubly atypical 
spectrum about any given vacuum has a well defined large $N$ limit, we may 
conclude that the states in this limiting spectrum (including the vacua 
themselves) are exact quantum states of M-theory on the pp-wave
(assuming the validity of the Matrix Theory conjecture).  

There are a number of interesting open questions and directions for 
future work. 

One question is whether any of the atypical multiplets for the 
single-membrane vacuum receive an energy shift. We showed that this is 
prohibited by 
group theory for $SU(2)$, but not for $SU(N)$. On the other hand, in limited 
perturbative calculations, we did not find any atypical multiplet which received 
an energy shift, even in cases where the complementary atypical multiplet 
existed in the spectrum. It would be interesting to see if this holds
beyond the leading order in perturbation theory and if so, to
understand the underlying reason why these states do not combine. A
related question is whether a state that is protected in the $SU(N_1)$
theory for some $N_1$ is also protected in the $SU(N)$ theory for
higher $N$. We have seen some evidence for this through leading order
perturbative calculations, but it would be interesting to understand
whether or not this is true in general.

We have found that all states about the single membrane vacuum have
energy shifts at leading order in perturbation theory which are finite
in the large $N$ limit, extending our results in \cite{dsv} for the
states $(a^\dagger_{00})^n |0 \rangle$. A natural question is whether
this finiteness of the perturbative corrections holds also to higher
orders in perturbation theory. For example, one might evaluate the
second energy correction for the lowest energy typical state
$(a^\dagger_{00})^2 | 0 \rangle$. It would be quite remarkable if the
complete perturbative expansion were finite term-by-term for $N \to
\infty$, since the supermembrane field theory is superficially
quite non-renormalizable. Also, we showed in \cite{dsv} that the
height of the energy barrier between various vacua goes to zero in
this limit. If the perturbation theory does turn out to be finite, we
could use it to obtain reliable dynamical information about states in 
M-theory which are not protected by supersymmetry. 

Finally, it would be interesting to understand the representation theory 
for the superalgebras corresponding to various string theories on 
other pp-wave backgrounds and see whether 
similar interesting protected multiplets exist in those cases. A more
direct application of the analysis of this paper might be to the
ordinary AdS/CFT conjecture, since the superalgebra we studied in this
paper is a subalgebra of the $SU(2,2|4)$ superconformal algebra
governing type IIB string theory on $AdS^5 \times S^5$.

\section*{Acknowledgements}
The research of K.D. is supported in part by a David and Lucile Packard 
Foundation Fellowship 2000-13856. The work of M. M. Sh-J., and M.V.R is 
supported in part by NSF grant PHY-9870115 and in part by funds from the 
Stanford Institute for Theoretical Physics.

\appendix

\section{Supersymmetry generators in terms of modes at $\mu=\infty$}

Explicit matrix expressions for the superalgebra generators 
were provided in \cite{dsv} (Appendix B). In particular, for the non-trivial
supercharge, we have
\beas
Q_{I \alpha} &=& \sqrt{R} \tr 
\left( (\Pi^a - i {\mu \over 6R} X^a)
{\sf g}^a_{IJ}
\epsilon_{\alpha \beta} \psi^{\dagger J \beta} - (\Pi^i + i {\mu \over 3R} 
X^i)
 \sigma^i_\alpha {}^\beta \psi_{I \beta} \right.\\
& & \left. + {1 \over 2} [X^i, X^j] \epsilon^{ijk} \sigma^k_\alpha {}^\beta
\psi_{I \beta} - {i \over 2} [X^a, X^b] ({\sf g}^{ab})_I^J \psi_{J \alpha} +i
[X^i, X^a] {\sigma^i_{\ \alpha}}^{\ \lambda}\ {\sf g}^a_{IJ} 
\epsilon_{\lambda 
\beta} \psi^{\dagger J\beta} \right)
\eeas
where ${\sf g}^{ab} = {1 \over 2} ({\sf g}^a {\sf g}^{\dagger b} - {\sf g}^b
{\sf g}^{\dagger a})$
and the matrices ${\sf g}^a_{IJ}$ relate the antisymmetric product of 
two ${\bf 4}$ representations of $SU(4)$ to the vector of $SO(6)$, with 
\[
{\sf g}^a ({\sf g}^b)^\dagger + {\sf g}^b ({\sf g}^a)^\dagger = 2
\delta^{ab}\ .
\]   
Expanding $X$'s about the vacuum solutions, i.e.
\[
X^i={\mu\over 3R} J^i+ Y^i
\]
choosing the gauge $A_0=0$, and making the proper rescalings 
\[
Y^i\rightarrow \sqrt{{R\over \mu}}Y^i, \ \
X^a\rightarrow \sqrt{{R\over \mu}}X^a, \ \
t\rightarrow {1\over \mu} t
\]
the supercharges $Q_{I\alpha}$ take the form
\be
Q_{I\alpha}= Q^0_{I\alpha}+ \left({R\over \mu}\right)^{3\over 2}Q^1_{I\alpha} 
\ee
where
\bea\label{Q0}
Q^0_{I\alpha}&=&\sqrt{\mu}\tr\bigg\{ 
\left((\dot{X}^a - {i \over 6} X^a)
{\bf 1}+{i\over 3}[J^i, X^a]\sigma^i\right)_{\alpha}^{\beta} 
{\sf g}^a_{IJ}\epsilon_{\beta \lambda} \psi^{\dagger J \lambda}\\
\ \ \ & & \qquad  \qquad - \left(\dot{Y}^i + {i \over 3} Y^i
-{1\over 3}[J^i, Y^j]\epsilon^{ijk}\right){\sigma^i_{\ \alpha}}^{\ 
\beta}\psi_{I \beta}\bigg\}
\eea
and
\bea
Q^1_{I\alpha}=\sqrt{\mu}\tr\bigg\{ 
{1\over 2}[Y^i, Y^j]\epsilon^{ijk}{\sigma^i_{\ \alpha}}^{\ \beta}
\psi_{I \beta}-
{i\over 2}[X^a, X^b]{\sf g}_{ab\ I}^{\ J}\psi_{J \alpha}+
i[Y^i, X^a]{\sf g}^a_{\ IJ}(\sigma^i\epsilon)_{\alpha \beta} \psi^{\dagger 
J \beta}\bigg\}
\eea
In the $\mu=\infty$ limit, only the quadratic part $Q^0$ of the supercharge 
$Q$ remains. One may explicitly check that 
\beas
\{Q^{\dagger\ 0 I\alpha}, Q^0_{J\beta}\}&=&{\mu}\ 
\left( 2\delta^I_J\delta^\alpha_\beta H_2 -{1\over 3}\epsilon^{ijk}\sigma^{k\ 
\alpha}_{\ \beta}M_{ij}-{i\over 6}\delta^{\alpha}_{\ \beta} 
({\sf g}^{ab})^I_J M_{ab} \right)
\\
\{Q^{\dagger\ 0 I\alpha}, Q^1_{J\beta}\}+\{Q^{\dagger\ 1 I\alpha}, 
Q^0_{J\beta}\}
&=&{2\mu}\delta^I_J\delta^\alpha_\beta H_3 \\
\{Q^{\dagger\ 1 I\alpha}, Q^1_{J\beta}\}&=&{2\mu}\delta^I_J\delta^\alpha_\beta 
H_4 
\eeas

Now let us concentrate on the $Q^0$ piece. Expanding $X$'s in terms of the 
spherical harmonics and the eigen-modes about the irreducible $X=J$ vacuum 
we obtain \footnote{We are using the notations of section 5 of \cite{dsv}
except for $\chi^\dagger$, which in our present notation $\chi^\dagger_{jm}$
creates a state with $J^3=+m$.} 
\beas
Q_{I\alpha}&= &\sum_{j=1}^{N-1}
i\sqrt{{2\mu\over 3}} a_{j-1\ m}
\left(\matrix{
-\sqrt{j-m}\ {(\chi^{\dagger})_I}^{j-{1\over 2}\  m-{1\over 2}}  \cr
-\sqrt{j+m}\ {(\chi^{\dagger})_I}^{j-{1\over 2}\  m+{1\over 2}}  
}\right)_\alpha 
\\ \; \\
&+ & \sum_{j=1}^{N-1} i{\sqrt{{\mu\over 3}}} {\sf g}^a_{IJ} a^{a\ 
\dagger}_{jm}
\left(\matrix{
-\sqrt{j+m}\ (\chi^J)_{j-{1\over 2}\ -m-{1\over 2}}  \cr
-\sqrt{j-m}\ (\chi^J)_{j-{1\over 2}\ -m+{1\over 2}}  }\right)_\alpha
\\ \; \\
&+ &\sum_{j=0}^{N-1} i{\sqrt{{\mu\over 3}}} {\sf g}^a_{IJ} a^{a}_{jm}
\left(\matrix{
\sqrt{j-m+1}\ \eta^{\dagger\ J}_{j+{1\over 2}\ m-{1\over 2}}  \cr
-\sqrt{j+m+1}\ \eta^{\dagger\ J}_{j+{1\over 2}\ m+{1\over 2}}}\right)_\alpha
\\ \; \\
&+ &\sum_{j=0}^{N-1} i{\sqrt{{2\mu\over 3}}} 
b^{\dagger}_{j+1\ m}\left(\matrix{
\sqrt{j+m+1}\ \eta_I^{j+{1\over 2}\ -m-{1\over 2}}  \cr
-\sqrt{j-m+1}\ \eta_I^{j+{1\over 2}\ -m+{1\over 2}} }\right)_\alpha\ .
\eeas
In the above $a_{jm}, b_{jm}$ and $a^a_{jm}$ are the annihilation operators 
for the $\alpha_{jm}, \beta_{jm}$ and $ x^a_{jm}$ modes, respectively.
Note that sums for the last two lines are starting at $j=0$. The multiplet
generated through $j=0$, which is of course in the U(1) part of the matrix 
model, is the only one particle, doubly atypical representation about $X=J$ 
vacuum. 

For a general vacuum corresponding to $M_i$ coincident membranes at
radii $N_i$, we have a set of terms in $Q$ corresponding to each
ordered pair $(N_i, N_j)$. For a given $(N_i, N_j)$, the terms are
exactly as above, except that the creation and annihilation operators
are $M_i \times M_j$ and $M_j \times M_i$ matrices respectively, there
is an overall trace, and the sums range over the spins listed in Table 2.

\section{Perturbative calculations}

It was shown in \cite{dsv} that the energy shift for the $X=0$ vacuum 
state, i.e. the state $|0\rangle$ is zero (up to the second order in 
perturbation theory). Here we  explicitly 
perform the calculation of the energy shift 
for the lowest non-trivial excited states above the $X=0$ vacuum.
The lowest energy non-trivial states (i.e.
involving more than $U(1)$ oscillators), are 
\[   
|\psi_{ab} \rangle = {1 \over \sqrt{(N^2-1)}} \left(
\tr(A_a^\dagger A_b^\dagger)- 
{1\over N} \tr(A_a^\dagger) \tr(A_b^\dagger) \right) |0 \rangle\ .
\]   
This states have been normalized as 
\be\label{normal}
\langle\psi_{ab} |\psi_{cd} \rangle= 
\delta_{ac}\delta_{bd}+\delta_{ad}\delta_{bc}
\ee
and also they are chosen to be orthogonal to 
the states
\[
\tr(A_a^\dagger) \tr(A_b^\dagger)  |0 \rangle
\] 
which cannot receive any energy correction since they are in the
$U(1)$ part of the theory. Note that the states 21 states
$|\psi_{ab} \rangle$ are degenerate and their energy is ${\mu\over 3}$.
These 21 states can be put in $(1,1)$ and $(20,1)$ representations of
$SU(2)\times SU(4)$ corresponding to contrating the $SO(6)$ indices or
symmetrizing them and subtracting off the trace (we previously
computed the energy shift for the former state in \cite{dsv}.) 

In order to do perturbative calculations, first we note that the
cubic and quartic parts of the Hamiltonian expanded about $X=0$ vacuum are
\cite{dsv}
\beas
H_3 &=& \left({R \over \mu} \right)^{3 \over 2} \tr \left( \sqrt{3 \over 8} i
\epsilon^{ijk}(A_i A_j A_k + 3 A_i^\dagger A_j A_k + 3 A_i^\dagger A_j^\dagger
A_k + A_i^\dagger A_j^\dagger A_k^\dagger) \right.\\
&& \qquad - \sqrt{3 \over 2} \psi^{\dagger \alpha} \sigma^i_\alpha {}^\beta 
[A_i
+ A_i^\dagger, \psi_\beta] \\
&& \qquad \left. - {\sqrt{3} \over 2} \epsilon_{\alpha \beta} \psi^{\dagger
\alpha} {\sf g}^a [A_a + A_a^\dagger, \psi^{\dagger \beta}] + {\sqrt{3} \over
2} \epsilon^{\alpha \beta} \psi_\alpha {\sf g}^{\dagger a} [A_a + A_a^\dagger, 
\psi_\beta] \right)
\eeas  
and
\beas
H_4 =&& -\left({ R \over \mu}\right)^3 \tr \left( {1 \over 4} [X^i, X^j]^2 + 
{1
\over 2} [X^i, X^a]^2 + {1 \over 4} [X^a, X^b]^2  \right)
\eeas
It is easy to see that the energy shift at first order in perturbation theory
\[
\langle \psi_{ab} | H_3 | \psi_{cd} \rangle
\]
will be zero for all states, since all terms in $H_3$ have non-zero $H_2$
eigenvalue. Thus $H_3$ acting on any state gives a combination of basis state
all of whose energies are different than that of the original state.

At second order in perturbation theory, the energy shift is given by the 
eigenvalues of the $21\times 21$ matrix
\bea   
\Delta^{ab,cd} &=& \Delta_4^{ab,cd} + \Delta_3^{ab,cd} \nonumber\\
&=& \langle \psi_{ab} | H_4 | \psi_{cd} \rangle + 
\sum_n {1 \over E_\psi - E_n} 
\langle \psi_{ab} | H_3 | n \rangle \langle n | H_3 | \psi_{cd} \rangle 
\label{2pert0}
\eea

Let us first focus on the $\Delta_3^{ab,cd}$ part. We note that $H_3$ may be 
written as a sum of terms $H_3^i$ with specific $H_2$ eigenvalues,
\[ 
H_3 = \sum H^i_3
\]
such that
\[   
[H_2, H^i_3] = E_i H^i_3
\]
where $H_2$ is the quadratic Hamiltonian and all the $E_i$'s are distinct.

Using this decomposition, we find \cite{dsv}
\beas
\Delta_3^{ab,cd} &=& 
\sum_n {1 \over E_\psi - E_n} 
\langle \psi_{ab} | H_3 | n \rangle \langle n | H_3 | \psi_{cd} \rangle \\
&=& \sum_i -{1 \over E_i} \langle  \psi_{ab} |  (H^i_3)^\dagger H^i_3 | 
\psi_{cd} \rangle
\eeas
%
Since the vacuum energy shift was zero, we can ignore all ``disconnected''
contributions in which the creation operators from the initial state contract
only with annihilation operators from the final state since these terms will 
be
the same as for the vacuum shift. A further simplification arises from the 
fact
that the interaction vertices are written in terms of commutators of $X$'s 
with
the result that $\tr(A^\dagger)$ or $\tr(A)$ from the initial or final state
contracted with any of the interaction terms will vanish. Thus, we may ignore
the
\[   
- {1 \over N} \tr(A_a^\dagger) \tr(A_a^\dagger)
\] 
piece since it contributes only to the disconnected contributions.
Since the calculations are very much similar to the one appeared in 
\cite{dsv},  here we skip the details of the calculations and only show the 
result,
%
\[
\Delta_3^{ab,cd} = -216 N 
\left(\delta_{ac}\delta_{bd}+\delta_{ad}\delta_{bc}\right)\ .
\]
Now let us work out the $\Delta_4^{ab,cd}$ piece.
The connected part of $\Delta_4$ also receives two contributions, coming from
\[   
H_4^1 = - {1 \over 2} \tr(X^a X^b X^a X^b - X^a X^a X^b X^b)
\]   
and
\[
H_4^2 = - \tr(X^a X^i X^a X^i - X^a X^a X^i X^i)\ .
\]
%
Again using the tools introduced in \cite{dsv} we find
\beas
\Delta_4^{1\ ab,cd} &=& 
\langle \psi_{ab}| H_4^1 |\psi_{cd}\rangle \\
&=& {1 \over (N^2-1)} \cdot  {1 \over 2} \langle \tr(A_a A_b)
\tr(X^e X^e X^f X^f - X^e X^f X^e X^f) \tr(A_c^\dagger A_d^\dagger) \rangle\\
&=& 18N\ 
\left[9(\delta_{ac}\delta_{bd}+\delta_{ad}\delta_{bc})+2 
\delta_{ab}\delta_{cd}\right]
\eeas
and
\beas
\Delta_4^{2\ ab,cd} &=& \langle \psi_{ab}| H_4^2 |\psi_{cd}\rangle \\ 
& =& {1 \over (N^2-1)} \langle \tr(A_a A_b)
\tr(X^f X^f X^i X^i - X^f X^i X^f X^i) \tr(A_c^\dagger A_d^\dagger) \rangle\\
&=& 54N\ (\delta_{ac}\delta_{bd}+\delta_{ad}\delta_{bc})
\eeas
Putting the values if $\Delta_4$ and $\Delta_3$ together we find
\bea
\Delta^{ab,cd} &=& \Delta_3^{ab,cd} + \Delta_4^{1\ ab,cd} + \Delta_4^{2\ 
ab,cd}\cr
&=& 36 N \delta_{ab}\delta_{cd}\  \mu \left( {R \over \mu} \right)^3\ .
\eea
Since $\Delta^{ab,cd}$ is already diagonal we can easily read off the energy 
shifts for $(1,1)$ and $(20,1)$ states.
As we see the energy shift for $(20,1)$ is zero, while the energy shift for 
$(1,1)$ state is $36\times{36\over 12}N \mu \left( {R \over \mu}
\right)^3$, as we found in \cite{dsv}. 
The factor of ${1\over 12}$ comes from properly normalizing the 
singlet part of the state (\ref{normal}) which we have used.

\end{document}